\documentclass[twocolumn,aps,prc,tightenlines,floats,floatfix,nofootinbib]{revtex4}

\def\etal{{\it et.\ al.\/},$\,$}

\def\bea{\begin{eqnarray}}
\def\eea{\end{eqnarray}}
\def\etal{{\it et al.\/}}

\def\etal{{\it et al.\/}}
\def\sfrac#1#2{{\textstyle \frac{#1}{#2}}}

\newcommand{\bra}[1]{\langle #1|}
\newcommand{\ket}[1]{|#1\rangle}

\def\be{\begin{equation}}
\def\ee{\end{equation}}
\def\ba{\begin{eqnarray}}
\def\ea{\end{eqnarray}}

\usepackage{graphics}
\usepackage{graphicx}
\usepackage{epsf}
\usepackage{amsmath}
\usepackage{amssymb}

\begin{document}

\phantom{0}
\vspace{-0.2in}
\hspace{5.5in}
\parbox{1.5in}{ \leftline{JLAB-THY-09-1024}}

\vspace{-1in}

\title
{A relativistic quark model for the
$\Omega^-$ electromagnetic form factors}
\author{G. Ramalho$^{1,2}$, K. Tsushima$^{3}$ and
Franz Gross$^{1,4}$
\vspace{-0.1in}  }

\affiliation{
$^1$Thomas Jefferson National Accelerator Facility, Newport News,
VA 23606, USA \vspace{-0.15in}}
\affiliation{
$^2$Centro de F{\'\i}sica Te\'orica de Part{\'\i}culas,
Ave.\ Rovisco Pais, 1049-001 Lisboa, Portugal \vspace{-0.15in}}
\affiliation{
$^3$EBAC in Theory Center,
Thomas Jefferson National Accelerator Facility, Newport News,
VA 23606, USA \vspace{-0.15in} }
\affiliation{
$^4$College of William and Mary, Williamsburg, VA 23185, USA}

\vspace{0.2in}
\date{\today}

\phantom{0}

\begin{abstract}
We compute the $\Omega^-$ electromagnetic form factors
and the decuplet baryon magnetic moments using a quark model  application of the Covariant Spectator Theory.  Our predictions for the $\Omega^-$
electromagnetic form factors can be tested in the future
by lattice QCD simulations
at the physical strange quark mass.
\end{abstract}

\vspace*{0.9in}  
\maketitle

\section{Introduction}

The $\Omega^-$ baryon has a unique position in the baryon decuplet.
A naive SU(6) quark model describes $\Omega^-$ as
a state with
three strange quarks in a totally symmetric  flavor-spin-space, e.g. 
\begin{equation}
\ket{\Omega^-;S=\sfrac{3}{2},S_z=+\sfrac{3}{2}}= \ket{s {\uparrow}
s {\uparrow} s {\uparrow}}.
\label{Omegafswf}
\end{equation}
As the strange quark decays via the weak interaction, the $\Omega^-$
has an extremely long lifetime ($\tau \simeq 8\times 10^{-11}$ s)
compared to the other decuplet members which have at least one light quark.
Because of this, the  world's average of the measurements of the magnetic dipole moment has a high precision,
$\mu_{\Omega^-}= (-2.02\pm0.05)\mu_N$~\cite{Yao06},
with $\mu_N$ the nuclear magneton.

Long before its experimental determination, the
$\Omega^-$ magnetic moment was estimated using a SU(6) symmetric quark model
\cite{Beg64}, which gave $\mu_{\Omega^-} =-\mu_p=-2.79 \mu_N$
(where $\mu_p$ is the proton magnetic moment in units of nuclear magneton).
That model was improved considering
the individual contributions of the
quark magnetic moments and the SU(3) symmetry breaking
(naive or static quark model) leading to
$\mu_{\Omega^-} \simeq -1.8 \mu_N$~\cite{Hikasa92,Muller}.
The naive result was then corrected
including non-static corrections,
sea quark contributions,
quark orbital momentum effects,
relativistic effects and others, using several formalisms
\cite{Das80,Tomozawa82,Georgi83,Verma87,Krivoruchenko87,Kim89,Kunz89,Chao90}.
In 1991 the $\Omega^-$ magnetic
moment was measured at Fermilab~\cite{Deihl91}.
The result was $\mu_{\Omega^-}=(-1.94\pm0.22) \mu_N$.
The most recent measurement is from 1995 and gives
$\mu_{\Omega^-}= (-2.024\pm0.056)\mu_N$~\cite{Wallace95}.
The combination of the two results leads to
$\mu_{\Omega^-}= (-2.019\pm0.054)\mu_N$~\cite{Yao06,Wallace95}.
Several works followed with estimations of $\mu_{\Omega^-}$
\cite{Schwesinger92,Schlumpf93,Hong94,Barik95,Ha98,Linde98,Zhu98,Wagner00,Aliev00,Iqubal00,Kerbikov00,Franklin02,Sahu02,An06,Ledwig08}.
The $\Omega^-$ electric quadrupole moment
was also predicted
\cite{Gershtein81,Richard82,Isgur82,Leonard90,Krivoruchenko91,Butler94,Sahoo95,Wagner00,Aliev00,Buchmann02,Buchmann07,Aliev09,Geng09},
although there is no experimental result.
Several works estimate the $\Omega^-$ electromagnetic
radius \cite{Schwesinger92,Kunz89,Barik95,Sahoo95,Wagner00,Buchmann02,Buchmann07,Gobbi92}.
%
Also the $\Omega^-$ magnetic octupole moment
was been estimated \cite{Buchmann08,Aliev09,Geng09}.

The magnetic moment \cite{Bernard82,Leinweber92,Lee05,Aubin}
and the $\Omega^-$ form factors \cite{Leinweber92,Boinepalli09}
(including the electric quadrupole and magnetic dipole moments) have also been calculated using lattice QCD.
The study of the $\Omega^-$ mass
in lattice QCD is nowadays an important
topic of investigation
helping to constrain the (physical) strange quark mass
in the quenched and dynamical calculations
\cite{Lin08,Tiburzi08,Aoki09,Drach09}.

In this work we extend the quark model based on the Covariant Spectator Theory \cite{Gross,FixedAxis},  originally developed to describe the nucleon form factors 
and properties of the $\Delta$,
to the full decuplet of baryons containing strange quarks.
In the previous work this spectator formalism
was applied to the $\gamma N \to \Delta$ transition form
factors~\cite{NDelta,NDeltaD,LatticeD}
as well as the nucleon~\cite{Nucleon}
and $\Delta$~\cite{DeltaFF,DeltaDFF}
electromagnetic form factors.
The flavor-spin structure of the $\Omega^-$
is very similar to that of the $\Delta$.
One gets the $\Omega^-$ either by replacing the
$u$ quarks by the $s$ quarks in the $\Delta^{++}$,
or replacing the $d$ quarks by the $s$ quarks in the $\Delta^{-}$.

However, the  $\Delta$ is significantly more
unstable ($\tau_\Delta \simeq 6 \times 10^{-24}$ s) than $\Omega^-$,
and this makes it very hard to measure the $\Delta$
electromagnetic form factors
experimentally.
At present, to compare with theoretical predictions,
we usually have to rely on
the pseudo-data, namely those extracted from lattice QCD.
Even then one must be careful,
since the lattice QCD results are obtained with unphysical pion masses
(heavy quark masses) which induce extra ambiguities, as discussed in Refs.~\cite{DeltaFF,DeltaDFF}.
On the other hand, the situation for the $\Omega^{-}$ 
is completely different.
It is presently possible to extract the magnetic 
dipole moment~\cite{Lee05,Aubin}
and the electric and magnetic dipole form factors~\cite{Boinepalli09}
in lattice QCD with the physical strange quark mass
of $m_s \approx 100$ MeV. 
Thus, theoretical predictions of the $Q^2$ 
dependence for the $\Omega^-$ electromagnetic form factors can be
directly compared with the lattice QCD data.

\section{Spectator Quark Model}

The covariant spectator quark model that we are using  (developed
in Refs.~\cite{Nucleon,NDelta,NDeltaD}
for the SU(2) light quark sector)
describes spin 1/2 and 3/2 three-quark systems as states of an off-shell quark and an on-shell spectator diquark \cite{Nucleon,Gross06}.  The diquark is intended to be a simple representation of the two {\it noninteracting\/} on-shell spectator quarks,  with a mass that varies from $4m_q^2$ to infinity
($m_q$ is the quark mass).
To simplify the calculation while still preserving the important physics, the integral over this mass is evaluated at some mean value $m_D$ (this mass was previously denoted $m_s$, but in this work $m_s$ will be reserved for the strange quark mass), which becomes a parameter of the model.  As it turns out, this parameter scales out of all form factor integrals, so that the results are independent of it
and it does not enter into the fits \cite{NDelta,Nucleon}.
The vertex functions are symmetrized so that (in the  relativistic impulse approximation) all form factors and transition amplitudes become a sum over terms in which the photon couples to each flavor of (off-shell) quark in turn with the other two (on-shell) quarks composing the on-shell diquark.  In this way interactions with {\it all\/} of the quarks are counted without including couplings to the diquark (in fact, to include them would be to over count).  Finally, our quarks are {\it constituent\/} quarks, with a form factor of their own, modeled using vector meson dominance.

A decuplet baryon ($B$) with a spin 3/2 wave function
based on this quark-diquark model with a 
pure orbital S-state can be generically
written~\cite{NDelta,NDeltaD}:
\bea
\Psi'_B(P,k; \lambda, \lambda_B) &=&
-\psi_B(P,k)\left| B\right> \varepsilon_P^{\alpha \ast}
(\lambda) u_\alpha (P; \lambda_B)
\nonumber\\
&=& \Psi_B(P,k)\ket{B},
\label{eqPsiB1}
\eea
where $\Psi_B$ (which suppresses the polarizations 
$\lambda$ and $\lambda_B$ and extracts $\ket{B}$) 
is a shorthand notation we will use through this paper.
Here $P$ ($k$) is the baryon (diquark) momentum, $\lambda=0,\pm 1$ the
diquark polarization, $\lambda_B=0,\pm 1/2,\pm 3/2$
the baryon spin projection, $u_\alpha$ the Rarita-Schwinger vector spinor,
$\varepsilon_P^{\alpha \ast}$ the  polarization state of the outgoing diquark,
and $\left|B\right>$ is a flavor state which will be specified latter.
As for $\psi_B(P,k)$, it is
{\it a real scalar function} that models the
momentum distribution of the quark-diquark system.
For the diquark polarization states we adopt
the fixed-axis basis~\cite{FixedAxis}, where the diquark
spin states are characterized by the momentum of the baryon $P$,
instead of the diquark momentum $k$.
Although this choice might be unconventional, it generalizes the
non relativistic structure for both
the spin 1/2 and spin 3/2 cases \cite{NDelta,Nucleon}.
In addition, the wave function $\Psi_B(P,k)$
satisfies 
the equation $(M_B- \not \! \!P) \Psi_B=0$,
where $M_B$ is the baryon mass
\cite{Nucleon,FixedAxis,NDelta,NDeltaD}.\footnote{The Fixed-Axis polarization basis also has the advantage
  of allowing a complete identification of the angular
  momentum components of the wave function.
  See Refs.~\cite{NDelta,NDeltaD} for details.}
Eq.~(\ref{eqPsiB1}) is the flavor generalization
of the $\Delta$ S-state wave function we have used
successfully in the past \cite{NDelta,DeltaFF}.
In this work we assume that the decuplet baryons can be approximated
as a quark-diquark in a spatial S-wave state.
Although there is strong evidence for the
presence of D-states in the $\Delta$,
the D-states admixtures are small~\cite{LatticeD}, and
the dominant form factors, such as the electric charge
and magnetic dipole moment, can be well described
without the D-state components~\cite{DeltaFF,DeltaDFF,DeltaDFF2}.

\begin{table*}[t]
\begin{minipage}{7.1in}
\begin{tabular}{l  c  c c c}
\hline
\hline
$B$   & &  $\ket{B}$ & & $\overline{j_{iB}}$     \\
\hline
\hline
$\Delta^-$  && $ddd=\ket{1,-1,0}_D\ket{d}$   && $\sfrac{1}{2} \left[f_{i+}-3 f_{i-} \right]$   \\[0.05in]
$\Delta^0$  && $\sfrac{1}{\sqrt{3}}\left[ddu + dud + udd  \right]=\sqrt{\sfrac23}\Big\{\left|1,-1,0\right>_D\left| u\right>+\sfrac1{\sqrt{2}}\left|1,0,0\right>_D\left| d\right>\Big\}$  &&
 $\sfrac{1}{2} \left[f_{i+}-f_{i-} \right]$ \\[0.1in]
$\Delta^+$  && $\sfrac{1}{\sqrt{3}}\left[uud + udu + duu  \right]=\sqrt{\sfrac23}\Big\{\left|1,1,0\right>_D\left| d\right>+\sfrac1{\sqrt{2}}\left|1,0,0\right>_D\left| u\right>\Big\}$  &&
 $\sfrac{1}{2} \left[3f_{i+}-f_{i-} \right]$ \\[0.05in]
$\Delta^{++}$  && $uuu=\ket{1,1,0}_D\ket{u}$  &&  $\sfrac{1}{2} \left[f_{i+}+3 f_{i-} \right]$  \\[0.05in]
\hline\\[-0.1in]
$\Sigma^{\ast -}$ &&
$\sfrac{1}{\sqrt{3}} \left[dds + dsd + sdd  \right]=\sqrt{\sfrac23}\Big\{\sfrac1{\sqrt{2}}\left|1,-1,0\right>_D\left| s\right>+\left|\sfrac12,-\sfrac12,-1\right>_D\left| d\right>\Big\}$  &&
 $\sfrac{1}{3} \left[f_{i+} - 3 f_{i-}- f_{i0} \right]$
\\[0.1in]
$\Sigma^{\ast 0}$ &&
$\sfrac{1}{\sqrt{6}} \left[uds + dus +usd +  sud + dsu + sdu  \right]=\sfrac1{\sqrt{3}}\Big\{\left|1,0,0\right>_D\left| s\right>+\left|\sfrac12,\sfrac12,-1\right>_D\left| d\right>+\left|\sfrac12,-\sfrac12,-1\right>_D\left| u\right>\Big\}$
 &&
 $\sfrac{1}{3} \left[f_{i+} - f_{i0} \right]$
\\[0.1in]
$\Sigma^{\ast +}$ &&  $\sfrac{1}{\sqrt{3}} \left[uus  + usu + suu  \right]=\sqrt{\sfrac23}\Big\{\sfrac1{\sqrt{2}}\left|1,1,0\right>_D\left| s\right>+\left|\sfrac12,\sfrac12,-1\right>_D\left| u\right>\Big\} $
  &&
 $\sfrac{1}{3} \left[f_{i+} + 3 f_{i-}- f_{i0} \right]$
\\[0.1in]
\hline\\[-0.1in]
$\Xi^{\ast -}$ && $\sfrac{1}{\sqrt{3}} \left[dss  + sds + ssd  \right]=\sqrt{\sfrac23}\Big\{\left|\sfrac12,-\sfrac12,-1\right>_D\left| s\right>+\sfrac1{\sqrt{2}}\left|0,0,-2\right>_D\left| d\right> \Big\}$   &&
 $\sfrac{1}{6} \left[f_{i+} - 3 f_{i-}- 4 f_{i0} \right]$
\\[0.1in]
$\Xi^{\ast 0}$ & $\;$& $\sfrac{1}{\sqrt{3}} \left[uss  + sus + ssu  \right]=\sqrt{\sfrac23}\Big\{\left|\sfrac12,\sfrac12,-1\right>_D\left| s\right>+\sfrac1{\sqrt{2}}\left|0,0,-2\right>_D\left| u\right>\Big\}$ &
&
 $\sfrac{1}{6} \left[f_{i+} + 3 f_{i-}- 4 f_{i0} \right]$
\\[0.1in]
\hline
$\Omega^-$     &&  $\left|sss\right>=\left|0,0,-2\right>_D\left|s\right>$  &&  $-f_{i0}$\\[0.02in]
\hline
\hline
\end{tabular}
\end{minipage}
\caption{Flavor wave functions $\left|B\right>$ expressed both in terms of their three-quark content and their quark-diquark content, using the diquark notation of Table \ref{tableII}.  The right column gives $\overline{j_{iB}}$.
One gets the contribution
of the electric charge function $\tilde e_B$ for $i=1$,
while for $i=2$ one gets the
contribution of the anomalous  magnetic moment function $\tilde \kappa_B$.}
\label{tableJi}
\end{table*}

\begin{table}
\begin{minipage}{3in}
\begin{tabular}{l c}
\hline
\hline
$\left|I,I_z,S\right>_D\qquad$ &  $\left|q_1q_2\right>$ \\
\hline
$\left|1,1,0\right>_D$ & $ \left|uu\right>$\\[0.05in]
$\left|1, 0, 0\right>_D$ & $\sfrac1{\sqrt{2}}\Big\{ \left|ud\right>+\left|du\right>\Big\}$\\[0.05in]
$\left|1,-1,0\right>_D$ & $ \left|dd\right>$\\[0.05in]
\hline
$\left|\sfrac12,\sfrac12,-1\right>_D$ & $\sfrac1{\sqrt{2}}\Big\{ \left|us\right>+\left|su\right>\Big\}$\\[0.1in]
$\left|\sfrac12,-\sfrac12,-1\right>_D\qquad$ & $\sfrac1{\sqrt{2}}\Big\{ \left|ds\right>+\left|sd\right>\Big\}$\\[0.1in]
\hline
$\left|0,0,-2\right>_D$ & $\left|ss\right>$\\[0.05in]
\hline
\hline
\end{tabular}
\end{minipage}
\caption{Diquark wave functions, with $I$ and $I_z$ the 
diquark isospin and its $z$ projection, and $S$ the diquark strangeness.}
\label{tableII}
\end{table}

\subsection{Electromagnetic current}

The electromagnetic current of the baryon
in a elastic process can be written in the covariant spectator quark model
\cite{Nucleon,NDelta,NDeltaD,DeltaDFF}:
\be
J_B^\mu(q)=
\sum_{a=1}^3 \sum_\lambda
\int_{k_a} \overline{\Psi^\prime}_B 
(P_+,k_a) j_a^\mu(q) \Psi_B^\prime (P_-,k_a),
\label{eqJ1}
\ee
where $P_-$ ($P_+$) is the initial (final) baryon momentum,  $k_a$ the momentum of the $a$-th diquark (the companion to the $a$-th quark, which has momentum $P-k_a$), and
$q=P_+-P_-$.
As for $j_a^\mu$ it represents the $a$-th quark current operator.
 Note that Eq.~(\ref{eqJ1}) corresponds to an impulse approximation in which the electromagnetic interaction is described as a sum over terms in which the photon couples to {\it each\/} of the three quarks in turn.
The integral sign $\int_k$ is a short-hand notation for
\bea
\int_k\equiv\int \frac{d^3 k}{(2\pi)^3 (2E_D)},
\eea
the covariant integration volume, where $E_D= \sqrt{m_D^2+k^2}$ is the energy of the
on-mass-shell diquark with mass $m_D$.
The sum over $a$ includes the interactions with all three quarks included in the $\left|B\right>$ state (as described further below), but since the wave function must be exactly symmetric in spin-flavor-coordinate space, each of the three terms in the sum is exactly identical, so that the current matrix element is simply
\be
J_B^\mu(q)=
3 \sum_\lambda
\int_{k} \overline{\Psi}_B (P_+,k)\bra{B} j^\mu(q)\ket{B} \Psi_B (P_-,k),
\label{eqJ2}
\ee
where, for definiteness, we take $k=k_3$ to be four-momentum of the third diquark (but the choice does not matter).

The quark electromagnetic current can be expressed
in terms of a Dirac $j_1$ and a Pauli $j_2$
form factors \cite{Nucleon,NDelta}:
\be
j^\mu (q)= j_1 (Q^2) \gamma^\mu
+ j_2(Q^2) \frac{i \sigma^{\mu \nu}q_\nu}{2M_N},
\label{eqJq}
\ee
where $M_N$ is the nucleon mass, and $Q^2=-q^2$.
The Dirac and Pauli form factors $j_i$ ($i=1,2$)
are diagonal operators in the 3$\times$3 flavor space which can be written,
\be
j_i (Q^2)=\sfrac{1}{6} f_{i +} (Q^2) \lambda_0 +
 \sfrac{1}{2} f_{i -}(Q^2) \lambda_3 
+ \sfrac{1}{6} f_{i 0}(Q^2) \lambda_s,
\label{eqJiq}
\ee
where $f_{i\,n}(Q^2)$, with $n=\pm,0$,
represents respectively the isoscalar ($+$),
isovector ($-$) and $s$ quark ($0$) form factors, and
$\lambda_0$, 
$\lambda_3$, and $\lambda_s$ are 
the diagonal matrices
\ba
&\lambda_0=\left(\begin{array}{ccc} 1&0 &0\cr 0 & 1 & 0 \cr
0 & 0 & 0 \cr
\end{array}\right), \hspace{.3cm}
&\lambda_3=\left(\begin{array}{ccc} 1& 0 &0\cr 0 & -1 & 0 \cr
0 & 0 & 0 \cr
\end{array}\right),
\label{eqL1L3} \\
&\lambda_s \equiv \left(\begin{array}{ccc} 0&0 &0\cr 0 & 0 & 0 \cr
0 & 0 & -2 \cr
\end{array}
\right),
\ea
that act on  
the quark wave function in flavor space 
\be
q=  \left(\begin{array}{c} u \cr d \cr
s \cr
\end{array}
\right).
\ee
Using the normalization $f_{1\, n}(0)=1$, we recover
the usual relation for the quark charge:
\be
j_1 (0)=
\sfrac{1}{2} \lambda_3 +
 \sfrac{1}{2\sqrt{3}}  \lambda_8,
\ee
with $\lambda_8=
\sfrac{1}{\sqrt{3}}\left[
\lambda_0 + \lambda_s\right]$,
 the SU(3) generator.
Equation~(\ref{eqJiq}) generalizes the
current in Refs.~\cite{Nucleon,NDelta}
to include $s$ quarks.
In Ref.~\cite{Nucleon}, $f_{2\pm}(Q^2)$ was normalized
to $f_{2\pm}(0) =\kappa_\pm$ in order to reproduce
the nucleon magnetic moments, $\mu_p$ and $\mu_n$.
This fixes the values, $\kappa_+= 1.639$ and $\kappa_-=1.825$.
The extension of this strange quarks
gives $e_s \kappa_s= \sfrac{1}{6} f_{20}(0)
\lambda_s$ 
which leads to the normalization $f_{20}(0)= \kappa_s$.


\subsection{Computing the flavor matrix elements of the current}

The flavor wave functions for the baryons in the decuplet are given in Table \ref{tableJi}.  They can be expressed in two ways: first, as a direct product of the flavor states of the three quarks, suitably symmetrized, or second, as a sum over direct products of a diquark state described by isospin, $I$, $z$ projection of the isospin, $I_z$, and strangness, $S$, times the appropriate flavor of quark number 3.  For example, the flavor wave function of the $\Sigma^{*0}$, the completely symmetric $uds$ (in this notation, particle 1 is a $u$ quark, particle  2 a $d$ quark, and particle 3 an $s$ quark) can be written in two equivalent forms
\ba
\ket{\Sigma^{*0}}&=&\sfrac{1}{\sqrt{6}} \Big[(du + ud)s +(ds  + sd)u+ (us + su)d  \Big]
\nonumber\\
&=&\sfrac1{\sqrt{3}}\Big\{\left|1,0,0\right>_D\left| s\right>+\left|\sfrac12,-\sfrac12,-1\right>_D\left| u\right>
\nonumber\\
&&\qquad+\left|\sfrac12,\sfrac12,-1\right>_D\left| d\right>\Big\}, \label{eq:10}
\ea
where the diquark states are defined in Table \ref{tableII}.  In the second line of Eq.~(\ref{eq:10}) we have written the state as a sum of terms with particles 1 and 2 treated as a diquark, and particle 3 as a single quark.  At this stage these two representations are completely equivalent, but later, when we convert these states to the covariant quark-diquark model, we allow quark 3 to be off-shell, and treat the diquark pair as a single particle of mass $m_D$, as described above.  This will break the symmetry between the three quarks, which is then restored by symmetrizing the state again.  When the electromagnetic matrix elements are calculated, these separate off-shell pieces do not interfere with each other, and the total matrix element is simply three times the matrix element with particle 3 off-shell (as discussed above).

Using these flavor wave functions we compute the flavor matrix elements of the current
\be
\overline {j_i}_B =3 \bra{B} j_i(3) \ket{B}, \qquad i=\{1,2\}
\label{eqjB}
\ee
where $\ket{B}$ is the flavor wave functions described above, and we have
incorporated the factor of 3 from Eq.~(\ref{eqJ2})
into the definition of  $\overline{j_i}_B$.
Using (\ref{eqjB}), the current (\ref{eqJ2})  becomes
\bea
J_B^\mu(q)&=&
\sum_\lambda
\int_{k} \overline{\Psi}_B (P_+,k)
\nonumber\\
&&\qquad\times
\left[
\overline{j_1}_{B} \gamma^\mu
+ \overline{j_2}_{B}
\frac{i \sigma^{\mu \nu} q_\nu}{2M_N}
\right] \Psi_B (P_-,k).\qquad
\label{eqJ3}
\eea
 
To calculate the matrix elements $\overline{j_i}_B$ 
we use the flavor wave functions from Table \ref{tableJi}.  
As an example, let us calculate $\overline{j_i}_B$
for $\Sigma^{\ast 0}$ state.
Using the quark-diquark representation
with the notation $B=\Sigma^{\ast 0}$, and recalling that
all of the diquark wave functions are orthonormal, we obtain three identical terms
\ba
\overline{j_i}_{\Sigma^{\ast 0}}&=& 3\bra{\Sigma^{\ast 0}}
\left\{
\sfrac{1}{6} f_{i+} \lambda_0 +
\sfrac{1}{2} f_{i-} \lambda_3 +
\sfrac{1}{6} f_{i0} \lambda_s
\right\} \ket{\Sigma^{\ast 0}}
 \nonumber \\
&= &\phantom{+}\bra{s}\left\{
\sfrac{1}{6} f_{i+} \lambda_0 +
\sfrac{1}{2} f_{i-} \lambda_3 +
\sfrac{1}{6} f_{i0} \lambda_s
\right\}\ket{s}
\nonumber\\
& &+\bra{d}\left\{
\sfrac{1}{6} f_{i+} \lambda_0 +
\sfrac{1}{2} f_{i-} \lambda_3 +
\sfrac{1}{6} f_{i0} \lambda_s
\right\}\ket{d}
\nonumber\\
& &+\bra{u}\left\{
\sfrac{1}{6} f_{i+} \lambda_0 +
\sfrac{1}{2} f_{i-} \lambda_3 +
\sfrac{1}{6} f_{i0} \lambda_s
\right\}\ket{u}
\nonumber\\
&=&-\sfrac{1}{3} f_{i0} +\left(
\sfrac{1}{6} f_{i+} - \sfrac{1}{2} f_{i-} \right)+
\left(\sfrac{1}{6} f_{i+} + \sfrac{1}{2} f_{i-} \right)
 \nonumber \\
&=&
\sfrac{1}{3} f_{i+} - \sfrac{1}{3} f_{i0}\, ,
\ea
where the diquark states appear in the second to forth lines only as a normalization factor of unity.
All of the matrix elements, calculated in the same manner, are given in the third column of Table \ref{tableJi}.

Similarly, we can evaluate the
electric charge $\tilde e_B$
and anomalous magnetic moment
$\tilde \kappa_B$ quark form factors defined by
\be
\tilde e_B(Q^2)= \overline{ j_1}_B(Q^2), \hspace{.4cm}
\tilde \kappa_B(Q^2)= \overline{ j_2}_B(Q^2).
\label{eqDefEB}
\ee
In the limit $Q^2=0$ the form factors give
the electric charge $e_B$ and anomalous magnetic moment
$\kappa_B$. 
%

\section{Baryon form factors}

The spin 3/2 baryon ($B$) electromagnetic form factors,
$F_i^\ast(Q^2)$ ($i=1,..4$),
are defined by the current
\cite{DeltaFF,Nozawa90,Pascalutsa07}:
\ba
J_B^\mu &=& - \bar u_\alpha
\left\{
\left[
F_1^\ast g^{\alpha \beta} +
F_3^\ast \frac{q^\alpha q^\beta}{4M_B^2}
\right] \gamma^\mu \right\}
u_\beta 
\nonumber \\
& & - \bar u_\alpha
\left\{
\left[
F_2^\ast g^{\alpha \beta} +
F_4^\ast \frac{q^\alpha q^\beta}{4M_B^2}
\right]
\frac{i \sigma^{\mu \nu}q_\nu}{2M_B}
\right\}
u_\beta.
\label{eqJB1}
\ea
%
Using the wave function (\ref{eqPsiB1}), the
hadronic current (\ref{eqJ2}) defined by the model,
and the generic structure of Eq.~(\ref{eqJB1}),
we can write the form factors
$F_i^\ast(Q^2)$ 
in terms
of the charge $\tilde e_B$ and
anomalous magnetic moment $\tilde \kappa_B$ form factors
defined in Eq.~(\ref{eqDefEB}).
Using the multipole form factors
given by a linear combination
of $F_i^\ast$~\cite{DeltaFF,Nozawa90,Pascalutsa07},
we get the expressions for the electric charge
and magnetic dipole form factors,
\ba
G_{E0}(Q^2)&=&
\left(
\tilde e_B(Q^2) - \tau \frac{M_B}{M_N} \tilde \kappa_B(Q^2) \right)
{\cal I}_B(Q^2),  
\label{eqGE0}
\\
G_{M1}(Q^2)&=&
\left(
\tilde e_B(Q^2) + \frac{M_B}{M_N} \tilde \kappa_B(Q^2) \right)
{\cal I}_B(Q^2),
\label{eqGM1}
\ea
where $\tau= \sfrac{Q^2}{4M_B^2}$.  Note that the  $Q^2=0$ limit of these form factors defines the charge [$e_B= G_{E0}(0)$] and magnetic dipole moment  
[$\mu_B= G_{M1}(0) \sfrac{e}{2M_B}$].  The factor ${\cal I}_B$ is
the overlap integral between the initial
and final scalar part  of the wave function in Eq.~(\ref{eqPsiB1}),
\be
{\cal I}_B(Q^2)=
\int_k \psi_B(P_+,k) \psi_B(P_-,k),
\label{eqIntB}
\ee
and is real.
In the limit $Q^2=0$, charge conservation requires ${\cal I}_B(0)=1$.

The derivation of Eqs.~(\ref{eqGE0}) and (\ref{eqGM1}) is given in Ref.~\cite{DeltaFF}
for the $\Delta$ case in the same S-state approximation.
The remaining form factors are
$G_{E2}$ and $G_{M3}$.
In the S-state approximation, $G_{E2}$ and $G_{M3}$
vanish \cite{DeltaFF}.
The differences between Eqs.~(\ref{eqGE0})-(\ref{eqGM1})
and the corresponding expressions in Ref.~\cite{DeltaFF}
are the baryon mass $M_B$ (which  replaces $M_\Delta$) and $\tilde e_B$ and $\tilde \kappa_B$ (which replace $\tilde e_\Delta$ and $\tilde \kappa_\Delta$).
%
Note that the mass ratio, $\sfrac{M_B}{M_N}$, results
from the simplification of the Pauli
current contribution of Eq.~(\ref{eqJq}),  possible because the states satisfy the Dirac equation, $(M_B - \not\! P) u_\alpha (P,\lambda_B)=0$ 
\cite{NDelta,NDeltaD,DeltaFF}.

\subsection{Baryon decuplet magnetic moments}

The $Q^2=0$ limit of $G_{M1}$ gives the baryon magnetic 
dipole moment $\mu_B$ in units of 
$\sfrac{e}{2 M_B}$
Converting the result into
nuclear magnetons $\mu_N$ gives
\be
\mu_B = G_{M1}(0) \frac{M_N}{M_B} \mu_N,
\label{eqGMb}
\ee
or
\be
\mu_B=
\left(
e_B + \frac{M_B}{M_N}\kappa_B \right) \frac{M_N}{M_B} \mu_N.
\label{eqMuB}
\ee
Recalling that $\kappa_B=\overline{j_2}_B(0)$ and
using the formulae for $\overline{j_2}_B$ from the third column of Table \ref{tableJi} with $f_{2+}(0)=\kappa_+= 2\kappa_u-\kappa_d$ and $f_{2-}(0)=\kappa_-= \frac23\kappa_u+\frac13\kappa_d$, the decuplet magnetic moments
can be expressed in terms of the anomalous moments of the three quarks (and their charges), as listed in table~\ref{tableMuB}.
If we ignore some pion cloud effects (discussed further below), the anomalous moments of the $u$ and $d$ quarks can be determined by a fit to the neutron and proton magnetic moments \cite{Nucleon}, giving $\kappa_u= 1.778$ and $\kappa_d=1.915$.
These values lead to the predictions \cite{DeltaFF}:
$\mu_{\Delta^{++}} =5.11 \mu_N$,
$\mu_{\Delta^{+}} =2.51 \mu_N$,
$\mu_{\Delta^{0}} =-0.09 \mu_N$
and  $\mu_{\Delta^{-}} =-2.70 \mu_N$.
For the other members of the decuplet results are dependent
on the strange quark anomalous magnetic moment, $\kappa_s$,
to be determined next.

\subsection{$\Omega^-$ magnetic moment}

For the $\Omega^-$ with $e_B=-1$
and $\kappa_B = -\kappa_s$
one gets
\be
\mu_{\Omega^-}=
-\left(
 1+ \frac{M_\Omega}{M_N}\kappa_s \right) \frac{M_N}{M_\Omega} \mu_N.
\label{eqMuOm}
\ee
For
a simple estimate of $\mu_{\Omega^-}$
we use $\kappa_s = \sfrac{1}{2}\left(\kappa_u+\kappa_d\right)$
corresponding to an approximate SU(3) limit,
since SU(2) is already broken ($\kappa_u \ne \kappa_d$).
This estimate, which we denote by SU$'$(3), gives  $\mu_{\Omega^-}= -2.41 \mu_N$,
where the experimental value is
$\mu_{\Omega^-}= (-2.02\pm0.05) \mu_N$,
which deviates by about 20\%.
The results for $\Sigma^\ast$ and $\Xi^\ast$
magnetic moments in the  SU$'$(3) approximation
are also given in table~\ref{tableMuB} together with the
nonrelativistic naive SU(6) quark model (NRQM) values
\cite{Hikasa92,Muller}.

Unfortunately there is no experimental data for the
other members of the decuplet
aside from the $\Delta$ (with no strange quarks).
Under these circumstances the experimental
value for $\mu_{\Omega^-}$ is the only
physical constraint available to fix $\kappa_s$.
Following the procedure used in Ref.~\cite{Nucleon} where $\kappa_u$ and $\kappa_d$ were constrained to fit  the nuclear magnetic moments
($\mu_p$ and $\mu_n$),
we use Eq.~(\ref{eqMuOm}) and adjust $\kappa_s$ to fit the $\Omega^-$ magnetic moment exactly.
This fixes $\kappa_s=1.462$.

Once $\kappa_s$ is fixed, we can make predictions
for all the strange decuplet baryon magnetic moments.
The results are presented in Table~\ref{tableMuB}
with the label CST for the quark model based on the Covariant Spectator Theory.

\begin{table}[t]
\begin{center}
\begin{tabular}{l c  c c r c r r r r}
\hline
\hline
$B$   &$\qquad$ &  $\kappa_B$ & & NRQM  &&   SU$'$(3)  && CST \\
\hline
$\Sigma^{\ast -}$ && $-\sfrac{1}{3}\left[2\kappa_d + \kappa_s\right]$ &&
$-$2.47 && $-2.57$   &&  -2.44 \\
$\Sigma^{\ast 0}$ &&
$\sfrac{1}{3}\left[2\kappa_u -\kappa_d-\kappa_s\right]$ && 0.32 && $-0.07$
&& 0.06\\
$\Sigma^{\ast +}$ && $\sfrac{1}{3}\left[4 \kappa_u -\kappa_s\right]$ &&
3.11 && 2.43 && 2.56 \\
\hline
$\Xi^{\ast -}$ &&
$-\sfrac{1}{3}\left[\kappa_d+2\kappa_s \right]$ && $-$2.11  && $-2.43$
&& -2.23 \\
$\Xi^{\ast 0}$ &&
$\sfrac{2}{3}\left[\kappa_u- \kappa_s \right]$ && 0.64 && $-0.05$
&& 0.21 \\
\hline
$\Omega^-$ && $-\kappa_s$ && $-$1.83 && $-$2.41 && -2.02 \\
\hline
\hline
\end{tabular}
\end{center}
\caption{Magnetic moments in nucleon magneton units $\mu_N= \sfrac{e}{2M_N}$,
where $M_N$ is the nucleon physical mass.
Note that $\mu_{\Omega^-}$ is not a prediction
because it was used to fix $\kappa_s$.}
\label{tableMuB}
\end{table}

\section{Model for the $s$ quark
current and wave functions}

In previous work~\cite{Nucleon,NDelta,NDeltaD}
the quark form factors were defined
in the  SU(2) sector ($u,d$) by using a parametrization for the isoscalar and isovector components  inspired by  vector meson dominance (VMD):
\ba
f_{1\pm}&=& \lambda +
(1-\lambda)
\frac{m_v^2}{m_v^2+Q^2}
+ c_\pm
\frac{M_h^2 Q^2}{\left(M_h^2+Q^2\right)^2},
\label{eqf1} \\
f_{2\pm}&=&
\kappa_\pm
\left\{ d_\pm
\frac{m_v^2}{m_v^2+Q^2}
+ (1-d_\pm) \frac{M_h^2}{M_h^2+Q^2}
\right\},
\label{eqFud}
\ea
where $m_v$ is the lightest vector meson mass
fixed to $m_v=m_\rho$ (for $I=1$) or $m_\omega$ (for $I=0$),
and $c_\pm$, $d_\pm$ are VMD coefficients adjusted to fit the nucleon form factors.  The model explicitly allows for the quarks to emerge as point-like particles at infinite $Q^2$ (as required by QCD) with an effective charge of $\lambda e_q$.  Fits to deep inelastic scattering (DIS) fixed $\lambda=1.21$ (for our most efficient model II).
The second term with the large mass $M_h$ is intended to approximate the sum over contributions from heavy vector mesons,
which characterize the short range structure of the VMD processes
important at high $Q^2$.
We take $M_h$ to be
twice the nucleon mass ($M_h= 2M_N$)
as in previous
work \cite{Nucleon,NDelta,NDeltaD,Lattice,LatticeD,DeltaFF,DeltaDFF}.

We extend the model to the strange sector
by defining the strange quark form factors:
\ba
f_{10}&=& \lambda +
(1-\lambda)
\frac{m_\phi^2}{m_\phi^2+Q^2}
+ c_0
\frac{M_h^2 Q^2}{\left(M_h^2+Q^2\right)^2},
\label{eqfs1} \\
f_{20}&=&
\kappa_s
\left\{ d_0
\frac{m_\phi^2}{m_\phi^2+Q^2}
+ (1-d_0) \frac{M_h^2}{M_h^2+Q^2}
\right\}.
\label{eqFs}
\ea
This parametrization adds two more parameters to the model,
($c_0$, $d_0$), in addition to $\kappa_s$.
Note that the $\phi$ meson is introduced
to model the strange quark sector
in the VMD framework.
In this framework the dressed electromagnetic interaction (in the $t=q^2$ channel) is described as a succession of $u\bar u$ and $d\bar d$ pairs interacting to generate the $\rho$ meson (for $I=1$) 
or the $\omega$ meson (for $I=0$), while for the strange quark sector
this succession of $s\bar s$ interactions forms the $\phi$ meson.

The baryon scalar functions, $\psi_B$,
that describe the momentum dependence of
the quark-diquark system, are parameterized in the  following way:
\ba
& &\psi_\Delta(P,k)= \frac{N_\Delta}{m_D(\alpha_1+ \chi_\Delta)^3}
\label{eqPsiDelta}\\
& &\psi_{\Sigma^*}(P,k)= \frac{N_{\Sigma^\ast}}{
m_D(\alpha_1+ \chi_\Sigma)^2(\alpha_2+ \chi_\Sigma)} \\
& &\psi_{\Xi^*}(P,k)= \frac{N_{\Xi^\ast}}{
m_D(\alpha_1+ \chi_\Xi)(\alpha_2+ \chi_\Xi)^2} \\
& &\psi_{\Omega}(P,k)= \frac{N_{\Omega}}{
m_D(\alpha_2+ \chi_\Omega)^3},
\ea
where
\ba
\chi_B&=& \frac{(M_B-m_D)^2-(P-k)^2}{m_D M_B}
\nonumber\\
&=&\frac{2P\cdot k}{m_DM_B}-2 \to 2\left(\sqrt{1+\frac{{\bf k}^2}{m^2_{_D}}}-1\right)
\nonumber\\
&\simeq& \frac{{\bf k}^2}{m^2_{_D}} \quad {\rm if}\: {\bf k}^2 <\!\!< m^2_{_D}  \, ,
\label{eqChi}
\ea
where the penultimate form holds in the rest frame of the baryon and shows that, in this frame, the argument $\chi_B$ is {\it independent\/} of the baryon mass.  We will therefore choose the parameters $\alpha_1$ and $\alpha_2$ to also be independent of the baryon mass.

Since the momentum distribution is {\it defined\/} in the baryon rest frame, we see that the wave functions of all the baryons are spherically symmetric and that $\alpha_i$ defines the momentum ranges of the
wave function in units of $k^2\sim m^2_D$. For example, in the rest frame the $\Omega$ wave function is
\bea
\psi_{\Omega}(P,k)&=& \frac{N_{\Omega}}{
m_D\Big(\alpha_2+2\left(\sqrt{1+\frac{{\bf k}^2}{m^2_{_D}}}-1\right)\Big)^3}
\nonumber\\
&\to& \frac{N_{\Omega}}{
m_D\Big(\alpha_2+ \frac{{\bf k}^2}{m^2_{_D}}\Big)^3} \quad {\rm if}\: {\bf k}^2 <\!\!< m^2_{_D} ,\qquad
\eea
showing that $\psi_{\Omega}$ behaves as a tripole at small $k=|{\bf k}|$, but only goes like $k^{-3}$ at large $k$.  However, when the bound state is moving in the $z$ direction with large $\eta=P_z/M_B$, the wave function is distorted and no longer spherical.  In a moving frame with $\eta\to\infty$, and ignoring all terms in the denominator of ${\cal O}(\eta^{-1})$,
\bea
\psi_{\Omega}(P,k)\to \frac{N_{\Omega}}{
m_D\Big(\alpha_2-2+2\eta\frac{k_-}{m_D}\Big)^3}
\eea
where $k_-=E_D-k_z$ is the minus light-cone component of the diquark momentum.  In coordinate space this wave function looks like a pancake.

With the inclusion of the factor $1/m_D$ in
the wave function definition, the diquark mass
dependence scales out of the integral Eq.~(\ref{eqIntB}),
and the final result becomes
independent of $m_D$~\cite{Nucleon}.
We note that the parameter $\alpha_1$ is associated with
the SU(2) sector ($u$ and $d$ quarks),
while $\alpha_2$ is associated with the $s$ quark.
A similar form to Eq.~(\ref{eqPsiDelta}) was introduced
in Ref.~\cite{NDelta} to describe the dominant
contribution of the $\gamma N \to \Delta$ transition
in a model based only on the S-wave components, as in this study.
(Originally the authors of Ref.~\cite{NDeltaD} chose a wave function dependent on two range parameters, but later on concluded that
one range parameter was enough,
particularly at small $Q^2$~\cite{LatticeD}.)

As $\kappa_s$ was fixed in the previous section
by the value of $\mu_{\Omega^-}$, only $c_0$, $d_0$ in
Eqs.~(\ref{eqFs}) and $\alpha_2$ in the
scalar functions $\psi_B(P,k)$ are not constrained.
The remaining parameters are already fixed
by the nucleon and $\Delta$ properties~\cite{Nucleon,NDeltaD}: the wave function parameter $\alpha_1$
was fixed in Refs.~\cite{NDeltaD};
the SU(2) current coefficients $c_\pm$, $d_\pm$
were adjusted in Ref.~\cite{Nucleon};
$\lambda=1.21$ from Ref.~\cite{Nucleon}.
Because there is no experimental data for
the electromagnetic form factors,
except for $\Omega^-$ magnetic moment,
we need to constrain the other parameters of our model using the recent lattice QCD data for
the baryon decuplet~\cite{Boinepalli09}.
In that work the $\Sigma^{\ast +}$, $\Sigma^{\ast -}$
and $\Xi^{\ast -}$ electromagnetic form factors
were estimated using quenched lattice QCD
at a momentum squared of $Q^2=0.230$ GeV$^2$,
for several values of $m_\pi$, in the range 300 MeV to
1 GeV~\cite{Boinepalli09}.
The work also estimated the $\Delta$ electromagnetic form factors.
However, because we want to use the same
wave function parametrization for the $\Delta$
as in previous work~\cite{NDeltaD,LatticeD,DeltaDFF},
we do not use the $\Delta$ data in Ref.~\cite{Boinepalli09}.
Study of the $\Delta$ electromagnetic form factors
with the inclusion of the D-states can be found
in Refs.~\cite{DeltaDFF,DeltaDFF2}.

\section{Using the lattice data}

To compare this model with
lattice QCD data we follow the procedure
presented in Refs.~\cite{Lattice,LatticeD}.  Briefly, since lattice calculations are normally carried out for a variety of light quark masses larger than their physical values (as reflected in pion masses heaver that the physical pion, usually for $m_\pi$ from about 300 MeV to about 1 GeV), we cannot compare our model to the lattice data unless we determine how our model will vary with the mass of the light quark.

To this end, we make the assumption that the range parameters in the baryon wave functions ($\alpha_1$ and $\alpha_2$) can be kept constant, but that the meson masses in the VDM description of the quark form factors ($m_v= m_\rho,\, m_\omega,$ or $m_\phi$, and $M_h= 2M_N$) will vary with the actual values obtained in the lattice calculations.  Furthermore, while the baryon wave functions do not depend on the baryon masses in their rest frame (justifying our assumption that  the range parameters are also independent of the baryon masses), they do depend on the masses in the moving frames encountered when the form factors are calculated at nonzero $Q^2$, and this is taken into account by using the lattice values of the baryon masses.
The dependence on $m_\rho\simeq m_\omega$ in lattice calculations is  parametrized by the simple analytic form \cite{Leinweber01}
\be
m_\rho= a_0 + a_2 m_\pi^2,
\ee
with $a_0=0.766$ GeV and $a_2=0.427$ GeV$^{-1}$, which are
consistent with the available quenched lattice QCD data.
As for the $\phi$ mass,
the parametrization in lattice QCD
requires some care.
Different from the SU(2) sector, the realistic
strange quark mass is currently used
in lattice QCD calculations.
In general, the proprieties of particles with strange quarks
are used to fix the strange quark mass on the lattice.
Although in dynamical simulations
sea quarks contribute for the $\phi$
meson mass~\cite{Lin08},
in the quenched simulation the $\phi$ mass
is independent of the light quark masses, thus
independent of the pion mass.
Because we apply our model to
the quenched lattice QCD data
determined at the physical strange
quark mass~\cite{Boinepalli09}, it may be justified to use
the physical $\phi$  mass, $m_\phi= 1019$ MeV.

The above procedure assumes that the valence
quark contributions are dominant,
and that the quenched QCD data simulates
well enough the valence quark effects.
The sea quark degrees of freedom
associated with intermediate meson states
are not considered explicitly.
In a formalism where the baryons are
the effective degrees of freedom
the virtual transitions between an 
initial baryon state and an intermediate
baryon plus meson state can also contribute
to the form factors.
In that case the photon can also
interact with the intermediate meson
adding extra contributions to the form factors.
According with $\chi$PT the light meson
gives the more important corrections.
It is known however that in lattice QCD calculations
the meson cloud effects are small in general
for $m_\pi >400$ MeV \cite{Detmold}.
It is also known that quenched QCD
underestimates the $\Delta^+$ magnetic moment
when the pion masses approaches to the physical point \cite{Young}.
That effect was indeed observed in Refs.~\cite{Lee05,Boinepalli09}.
However, the effects are not expected to be
dominant in the lattice
data analyzed in the present work\footnote{
  Considering the results of Ref.~\cite{Boinepalli09}
  for the $\Delta^+$ magnetic moment, and the difference
  between the $\mu_{\Delta^+}$ and the proton
  magnetic moment $\mu_p$ as a upper estimate
  of the pion cloud contribution, we conclude
  that these corrections are at most 30\%.}
since the magnitude of the meson cloud
becomes smaller when strange valence  quarks are present.


\begin{figure*}[t]
\vspace{.5cm}
\vspace{0.75cm}
\centerline{
\mbox{
\includegraphics[width=2.3in]{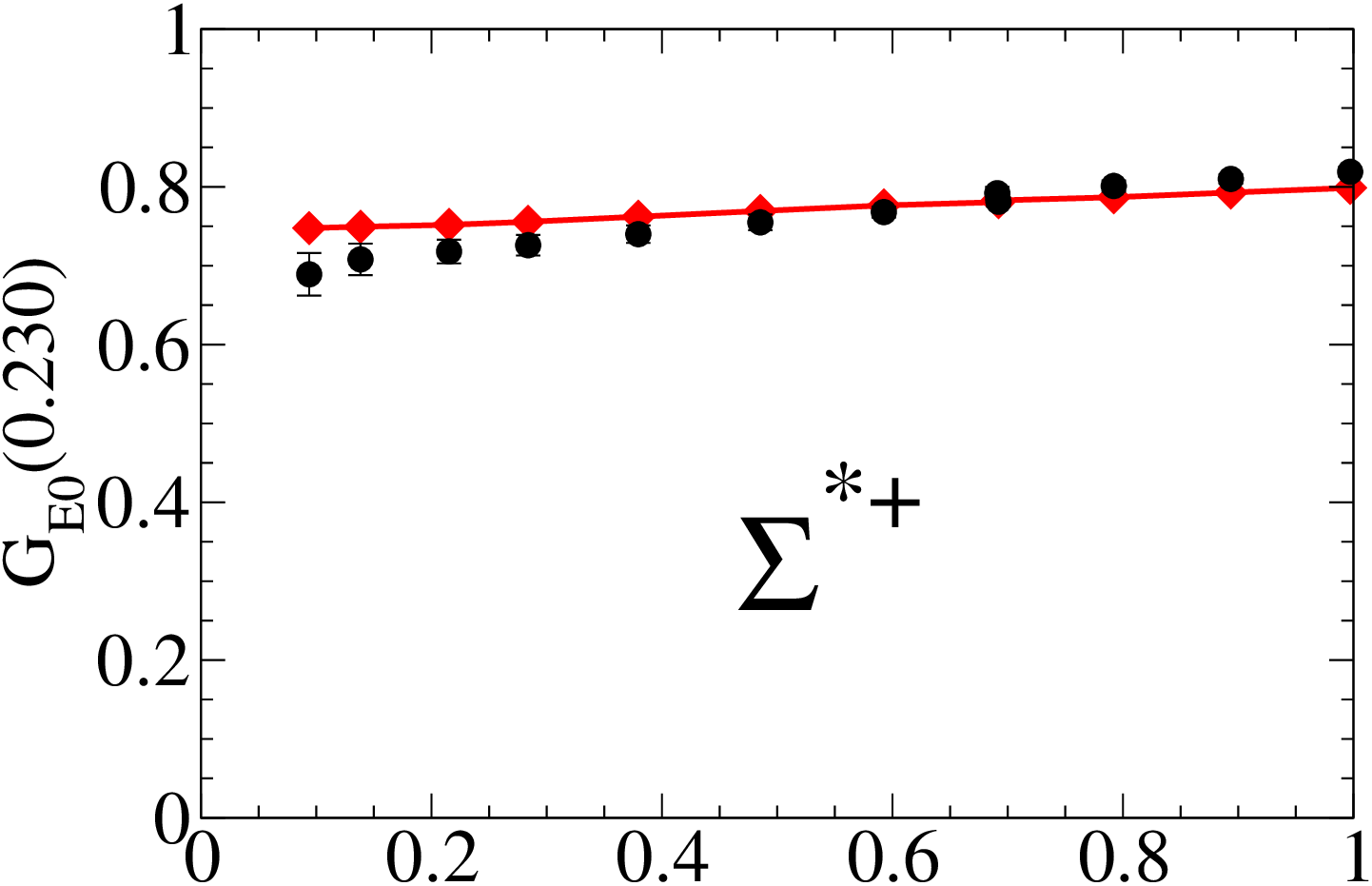} \hspace{.3cm}
\includegraphics[width=2.3in]{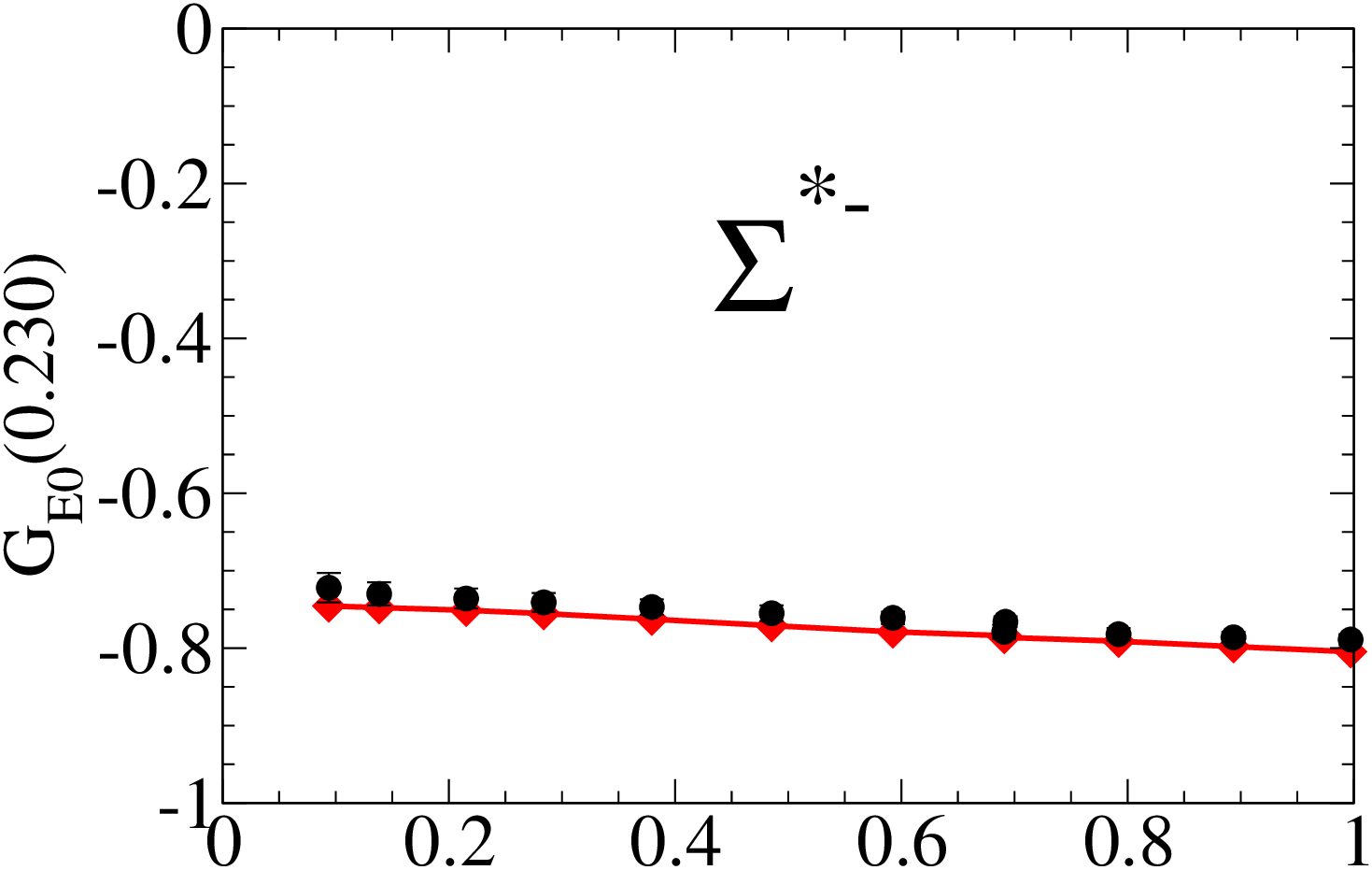} \hspace{.3cm}
\includegraphics[width=2.3in]{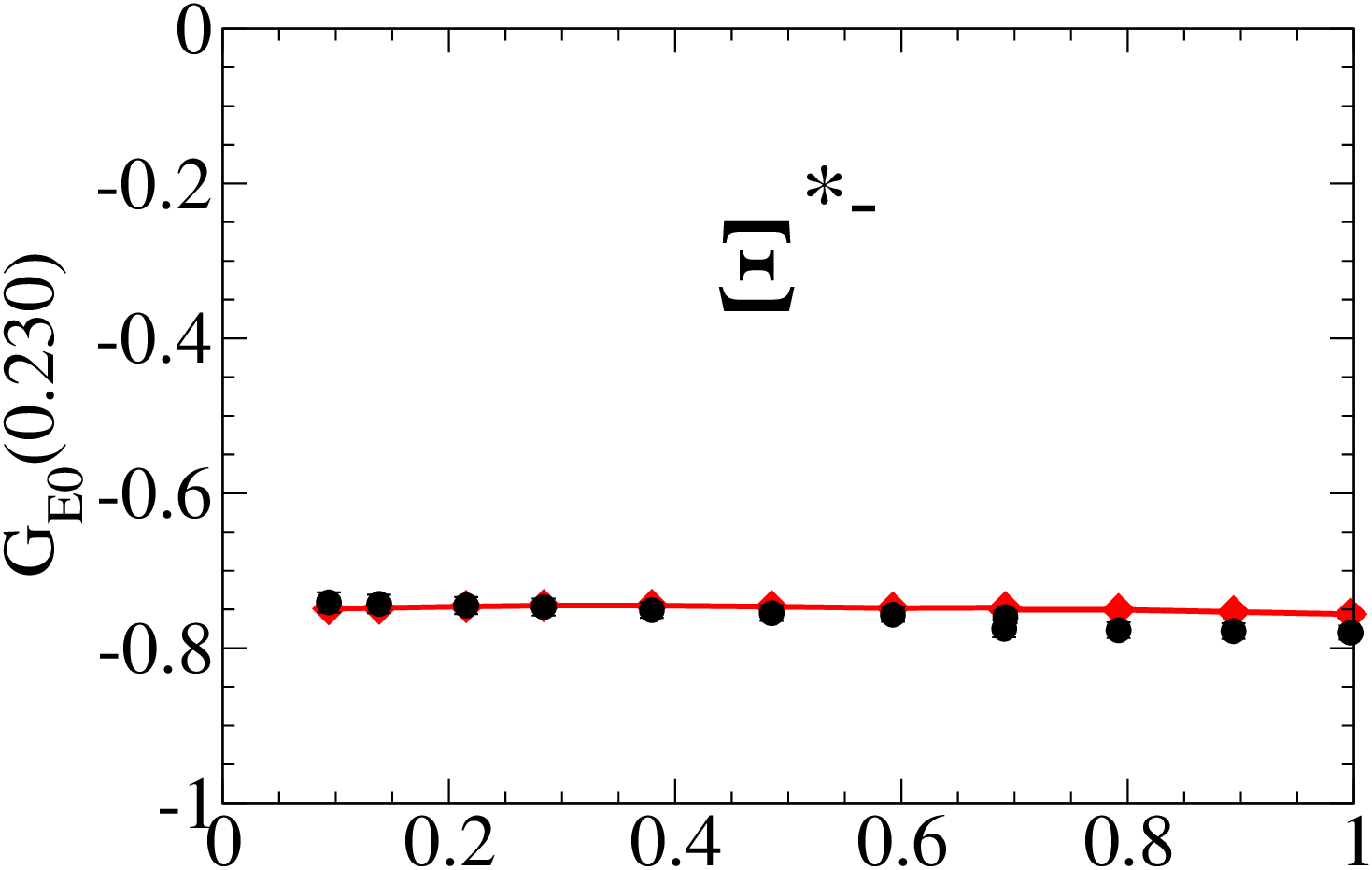}  \hspace{.3cm}
}}
\vspace{0.65cm}
\centerline{
\mbox{
\includegraphics[width=2.3in]{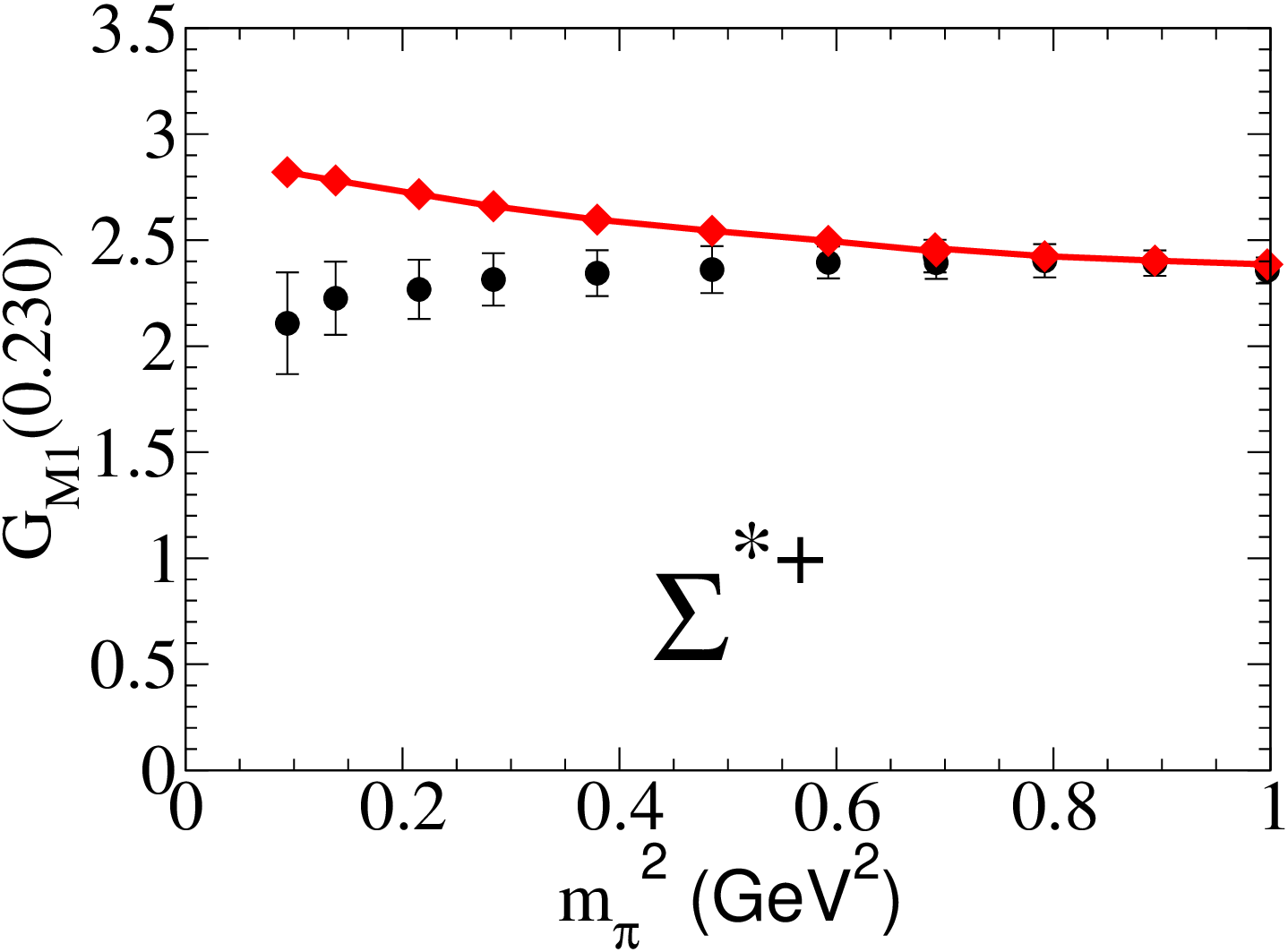} \hspace{.3cm}
\includegraphics[width=2.3in]{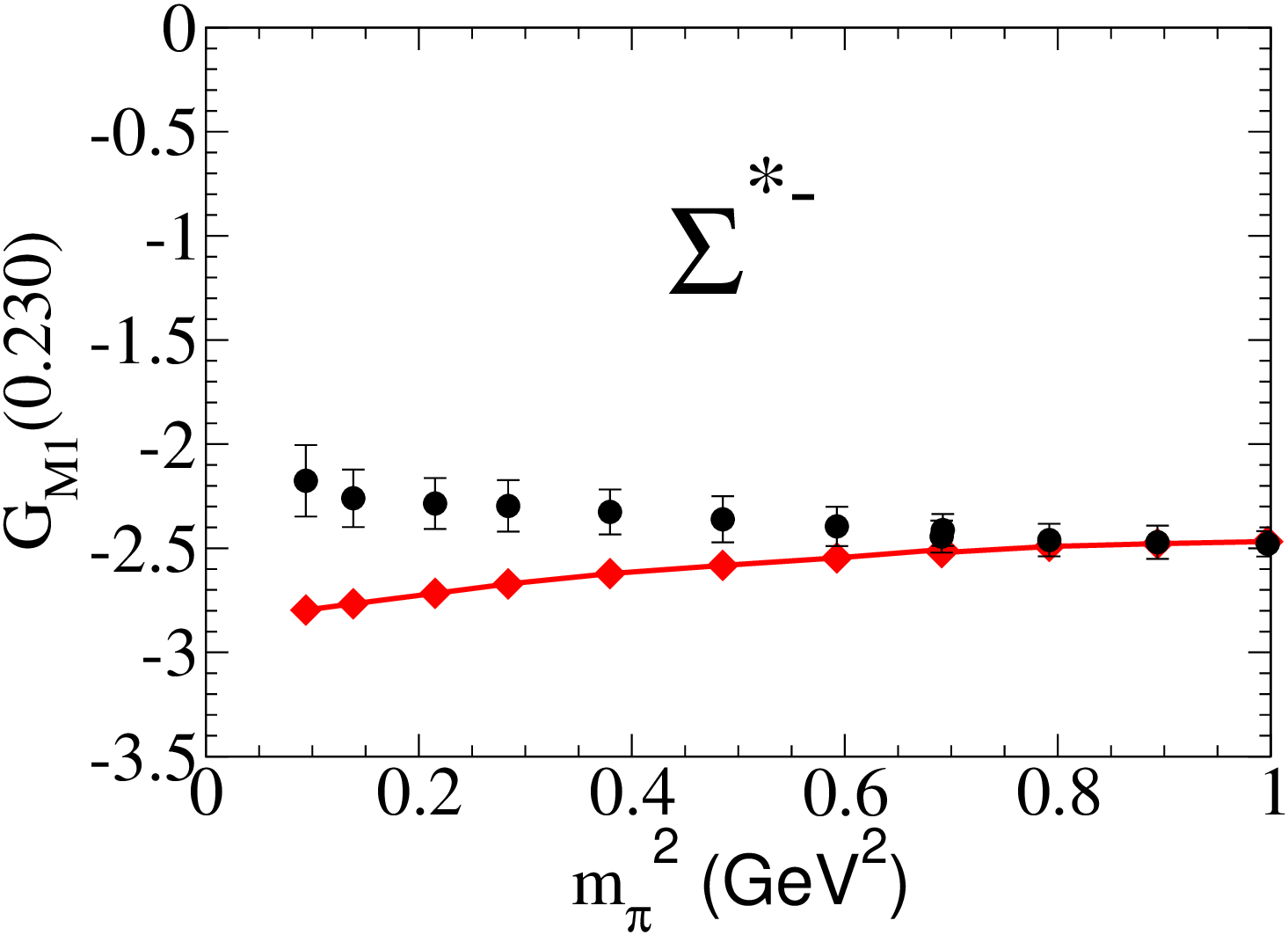} \hspace{.3cm}
\includegraphics[width=2.3in]{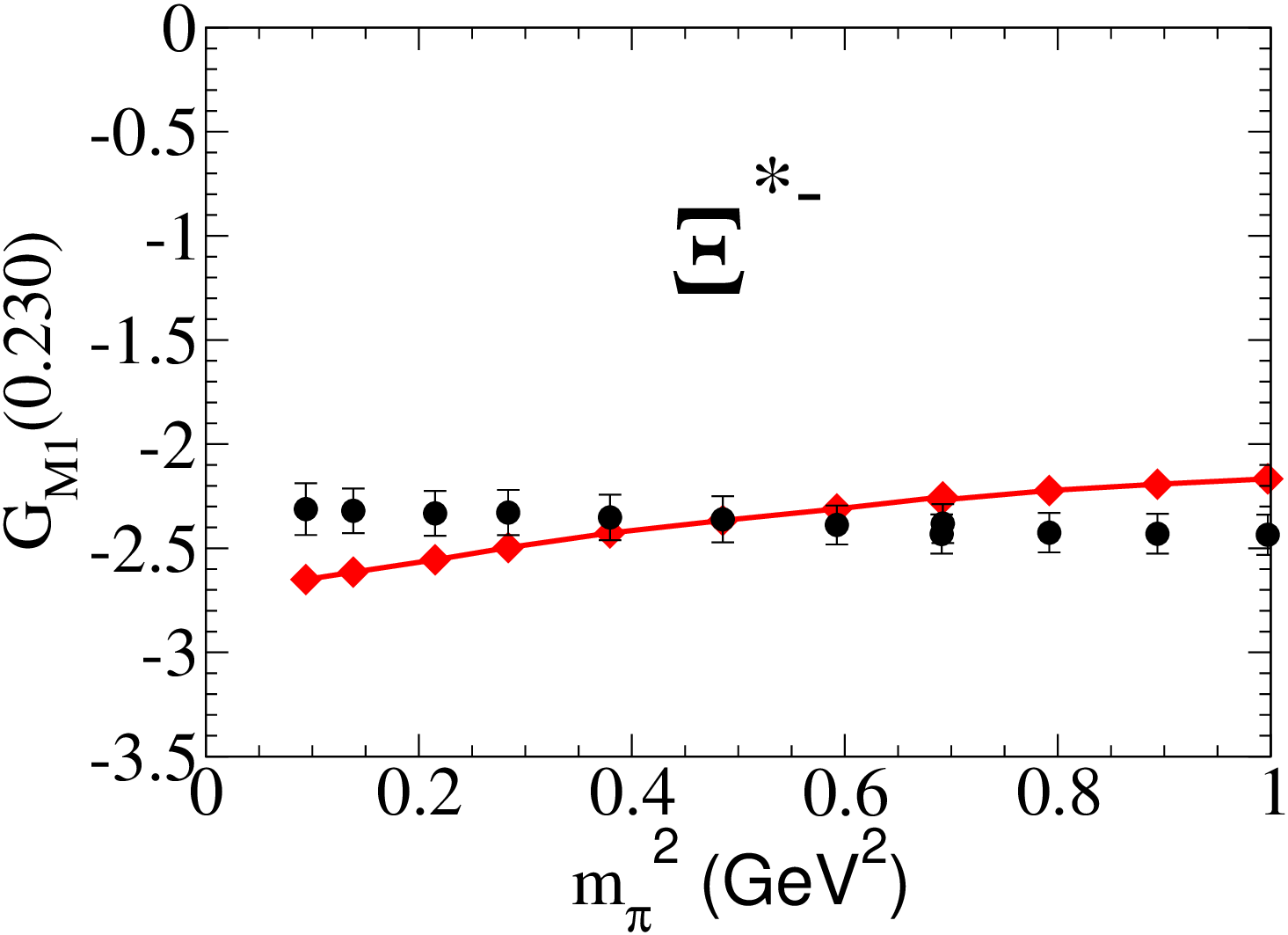} \hspace{.3cm}
}}
\caption{\footnotesize{
Results of our fit to the lattice data from
Ref.~\cite{Boinepalli09}.}}
\label{figFit}
\end{figure*}

\begin{table*}[t]
\begin{center}
\begin{tabular}{r cr rrr rr rr c cc cc}
\hline
\hline
$\alpha_1$, $\alpha_2$  &$\qquad$ &  $c_+$, $d_+$ &$\qquad$ & $c_-$, $d_-$ &  $\qquad$ &
$c_0$, $d_0$ & $\qquad$ & $\kappa_u$, $\kappa_d$ & $\qquad$ & $\kappa_s$ & $\qquad$ &
$N_\Delta$, $N_{\Sigma^\ast}$ & $\qquad$ &  $N_{\Xi^\ast}$, $N_{\Omega}$\\
\hline
0.3366  & & 4.160 & & 1.160  & & {\bf 4.427}  & & 1.777 & & {\bf 1.462} & &  2.594  & &  0.901  \\
{\bf 0.1630}  & &  -0.686 & & -0.686 & & {\bf -1.860}   & &   1.915 & & & &  1.553  & &  0.510  \\
\hline
\hline
\end{tabular}
\end{center}
\caption{Results from the fit to the lattice QCD data.
Only the variables in bold were adjusted
in this work; the others were previously determined from fits to the nucleon form factors and $\gamma N\to\Delta$ transitions.  The adjusted variables are:
$\kappa_s$, fixed by the experimental result for
$\mu_{\Omega^-}$ \cite{Wallace95};
$c_0$, $d_0$, and $\alpha_2$, fixed by the fit
to the lattice data \cite{Boinepalli09} shown in Fig.~\ref{figFit}.
The normalization factors are a consequence
of the values of $\alpha_1$ and $\alpha_2$.
}
\label{tableFit}
\end{table*}

In this work we use
the form factor data in Ref.~\cite{Boinepalli09}
and the lattice masses extracted for $\Sigma^\ast$ and $\Xi^\ast$.
For the nucleon we use the
nucleon mass derived from the same group,
with the same configuration in Ref.~\cite{Boinepalli06}.\footnote{
  For the pion mass $m_\pi^2=0.6920$(35) GeV$^2$
  there was no available $M_N$ mass in Ref.~\cite{Boinepalli06},
  so we took the result for $m_\pi^2=0.6910$(54) GeV$^2$.
  The difference between these $m_\pi^2$
  values is smaller than
  the statistical error-bars. }

To fix the strange quark mass in the quenched calculation, 
Ref.~\cite{Boinepalli09}  chooses the symmetry point where the light quark mass equals the strange quark mass.  At this point they find $m_K=m_\pi = 697$ MeV (and $m_\pi^2= 0.485$ GeV$^2$), which compares well with the experimental value of
$2m_K^2-m_\pi^2= (0.693$ GeV)$^2$, a constraint on the strange quark mass motivated by leading order chiral perturbation theory. 



\begin{figure*}[t]
\vspace{.5cm}
\centerline{
\mbox{
\includegraphics[width=2.2in]{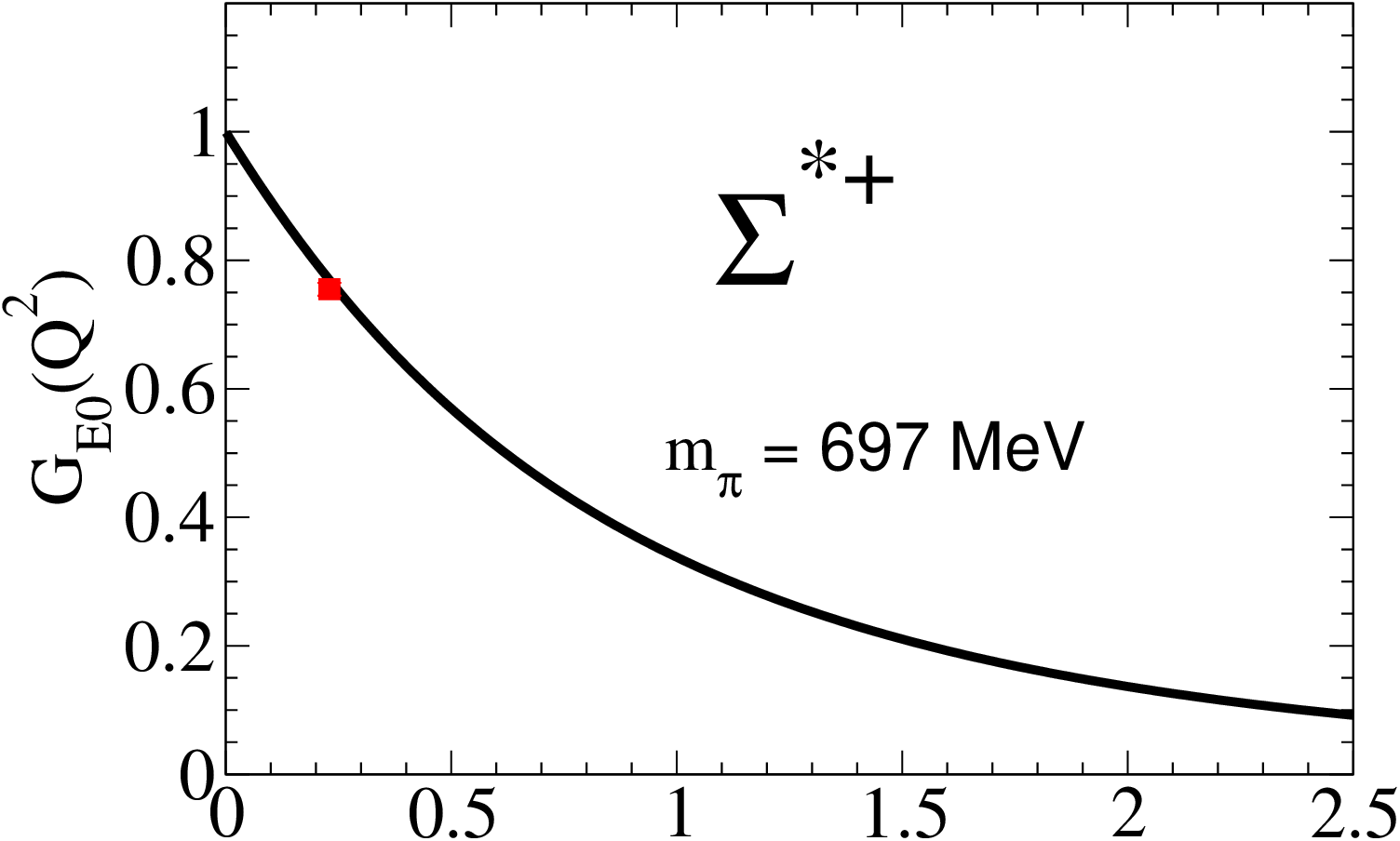}  \hspace{.1cm}
\includegraphics[width=2.2in]{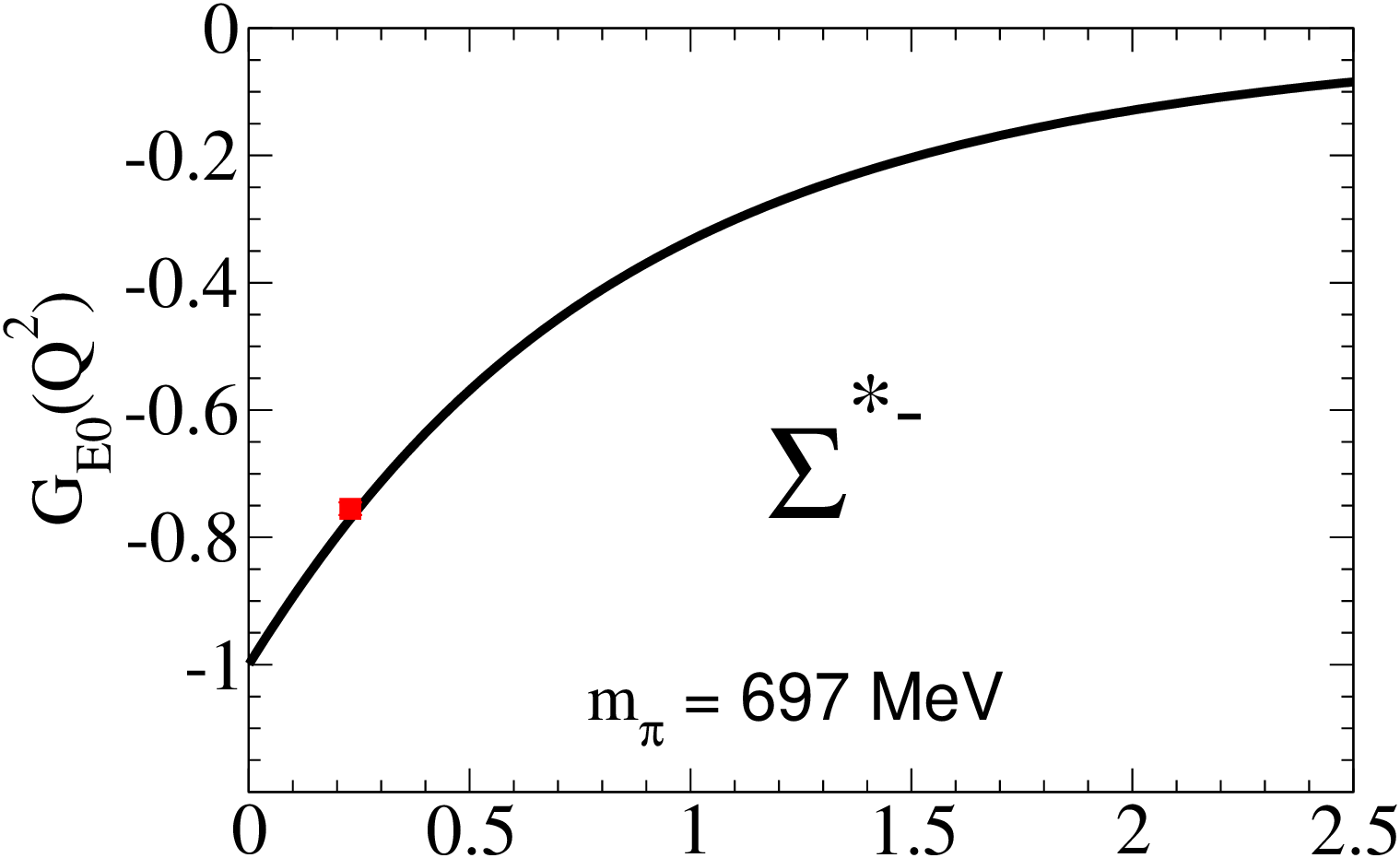}  \hspace{.1cm}
\includegraphics[width=2.2in]{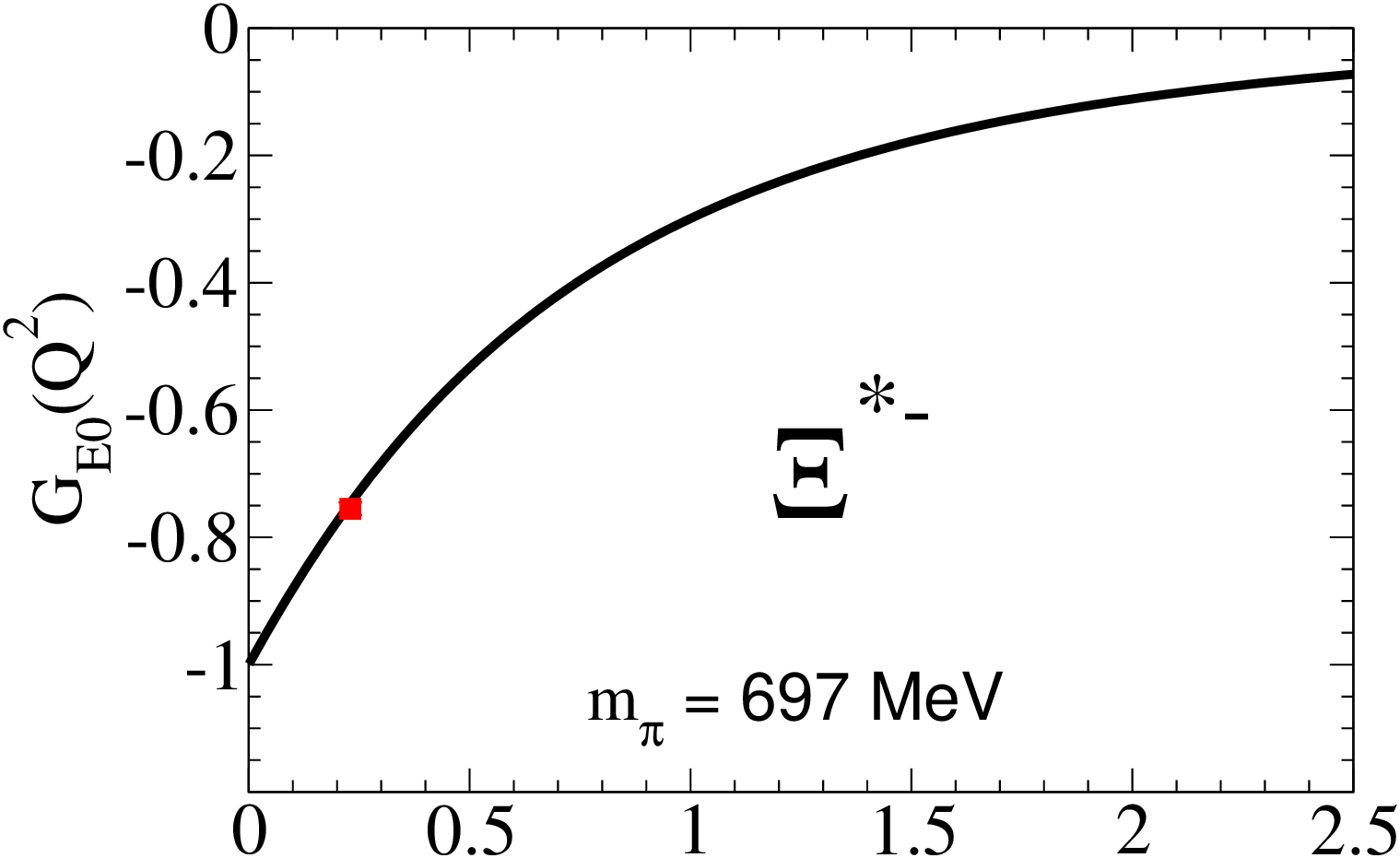}     \hspace{.1cm}
}}
\vspace{0.6cm}
\centerline{
\mbox{
\includegraphics[width=2.2in]{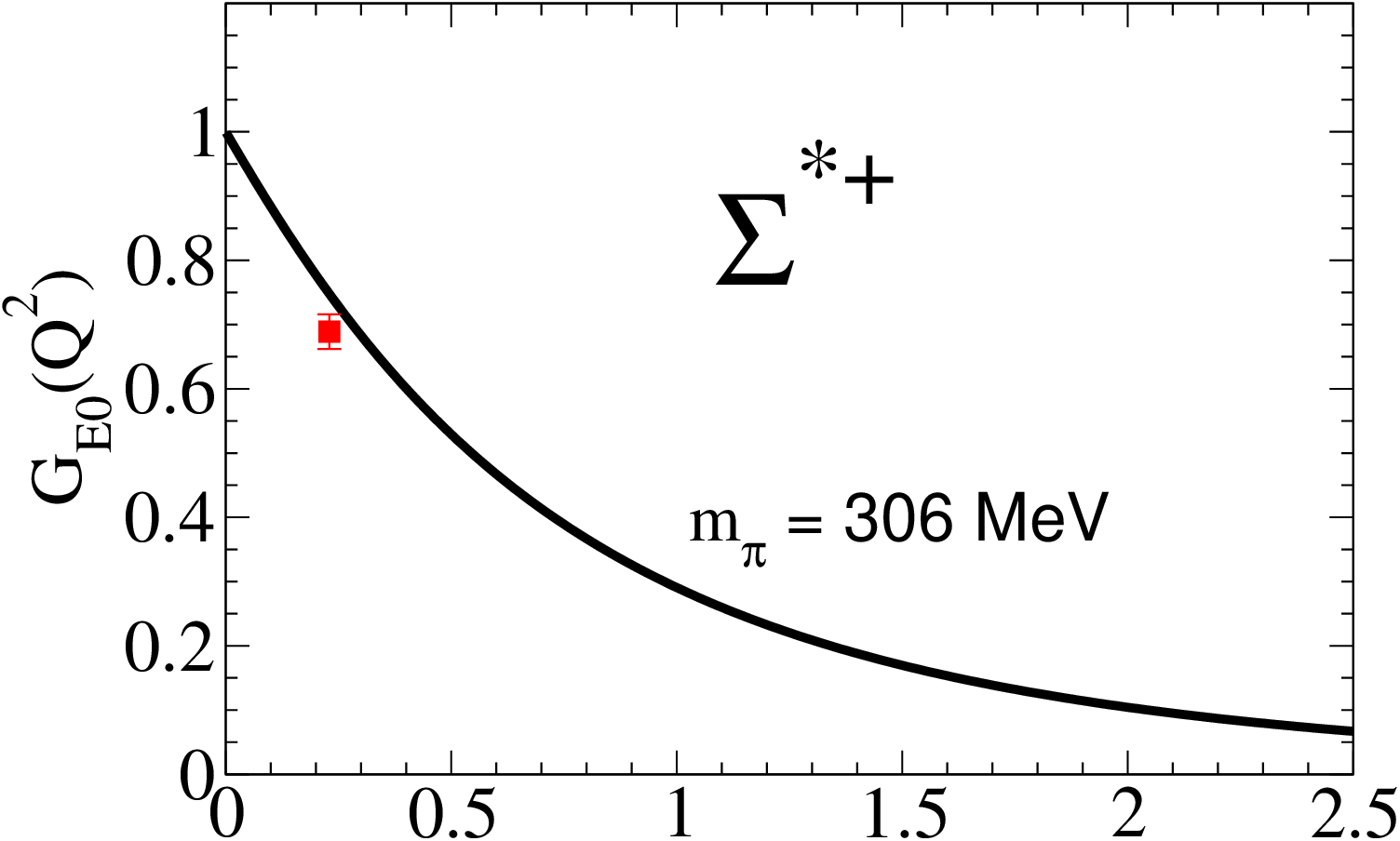}  \hspace{.1cm}
\includegraphics[width=2.2in]{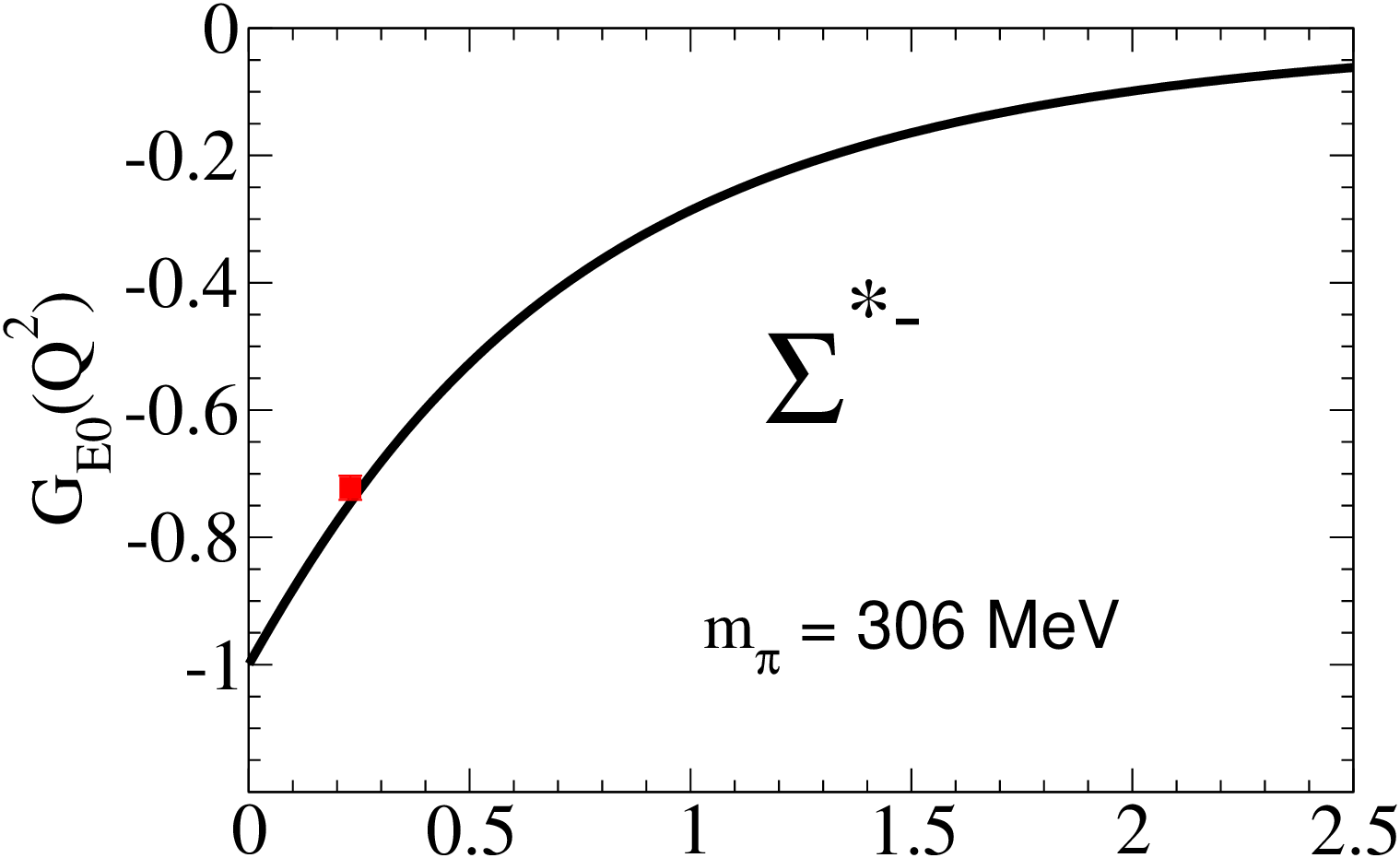}  \hspace{.1cm}
\includegraphics[width=2.2in]{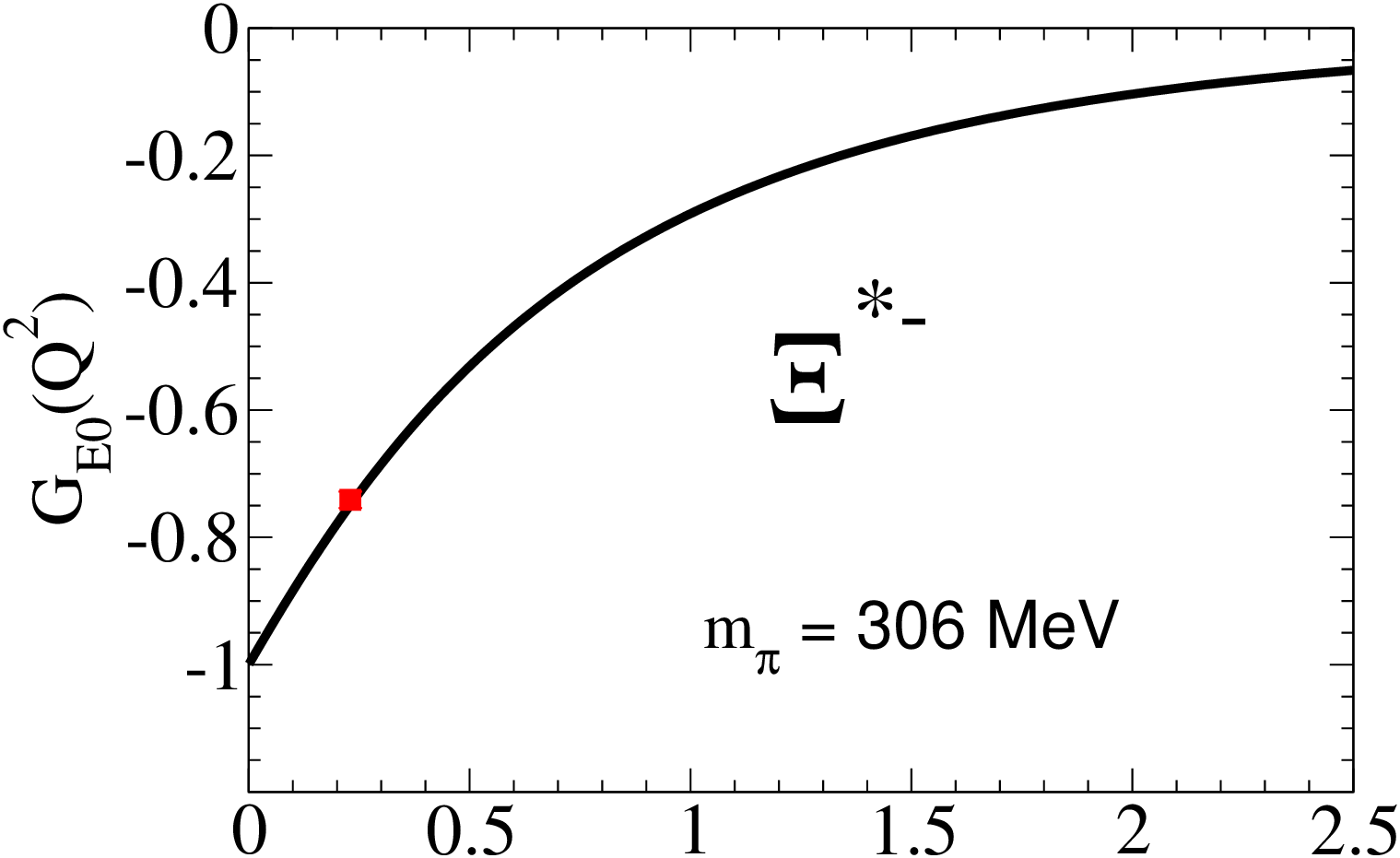}     \hspace{.1cm}
}}
\vspace{0.6cm}
\centerline{
\mbox{
\includegraphics[width=2.2in]{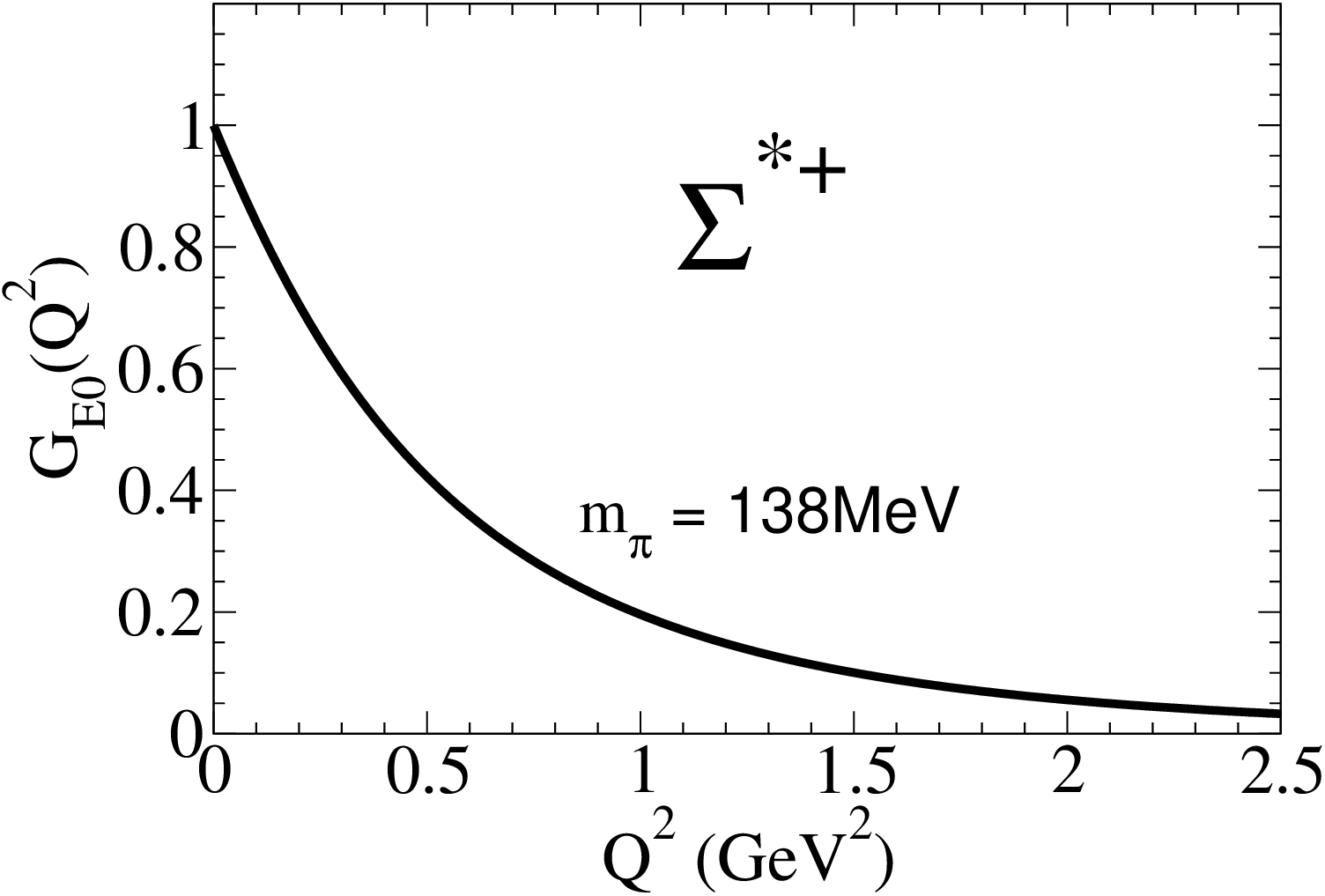}  \hspace{.1cm}
\includegraphics[width=2.2in]{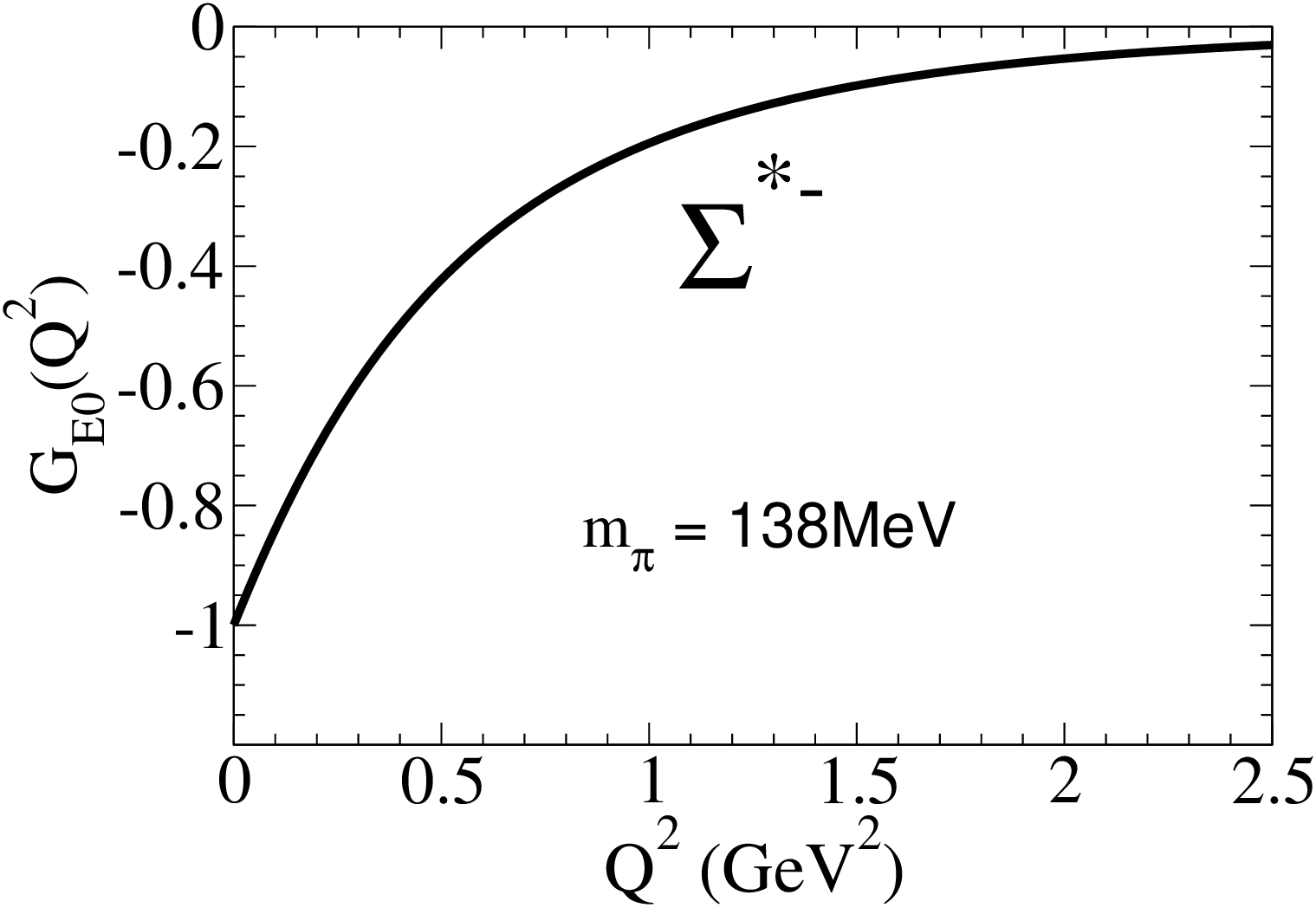}  \hspace{.1cm}
\includegraphics[width=2.2in]{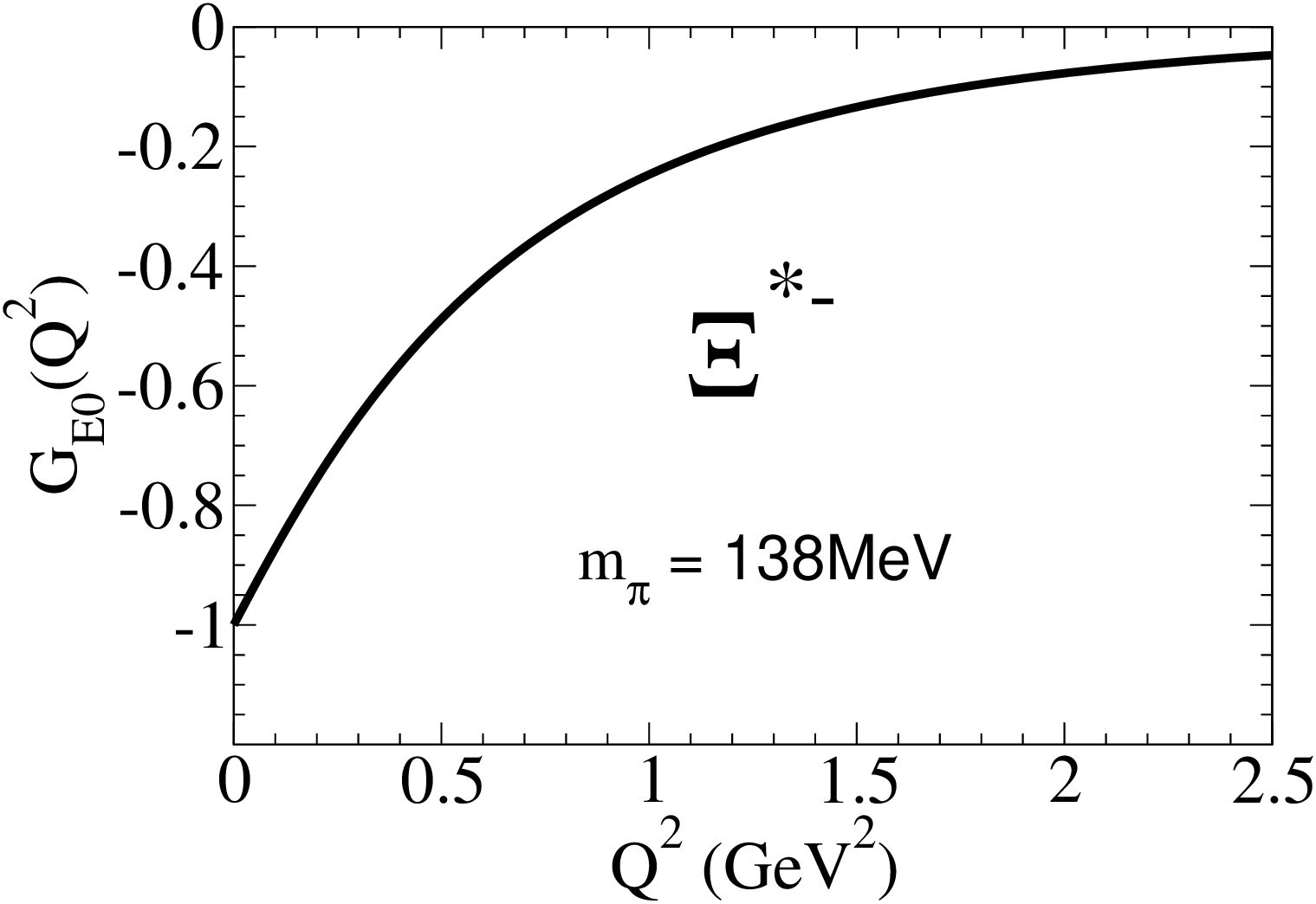}     \hspace{.1cm}
}}
\caption{\footnotesize{
$G_{E0}$ form factors for $\Sigma^{\ast +}$, $\Sigma^{\ast -}$
and     $\Xi^{\ast -}$. The quenched lattice QCD data is from
Ref.~\cite{Boinepalli09}.}}
\label{figGE0}
\end{figure*}

\begin{figure*}[t]
\vspace{.5cm}
\centerline{
\mbox{
\includegraphics[width=2.2in]{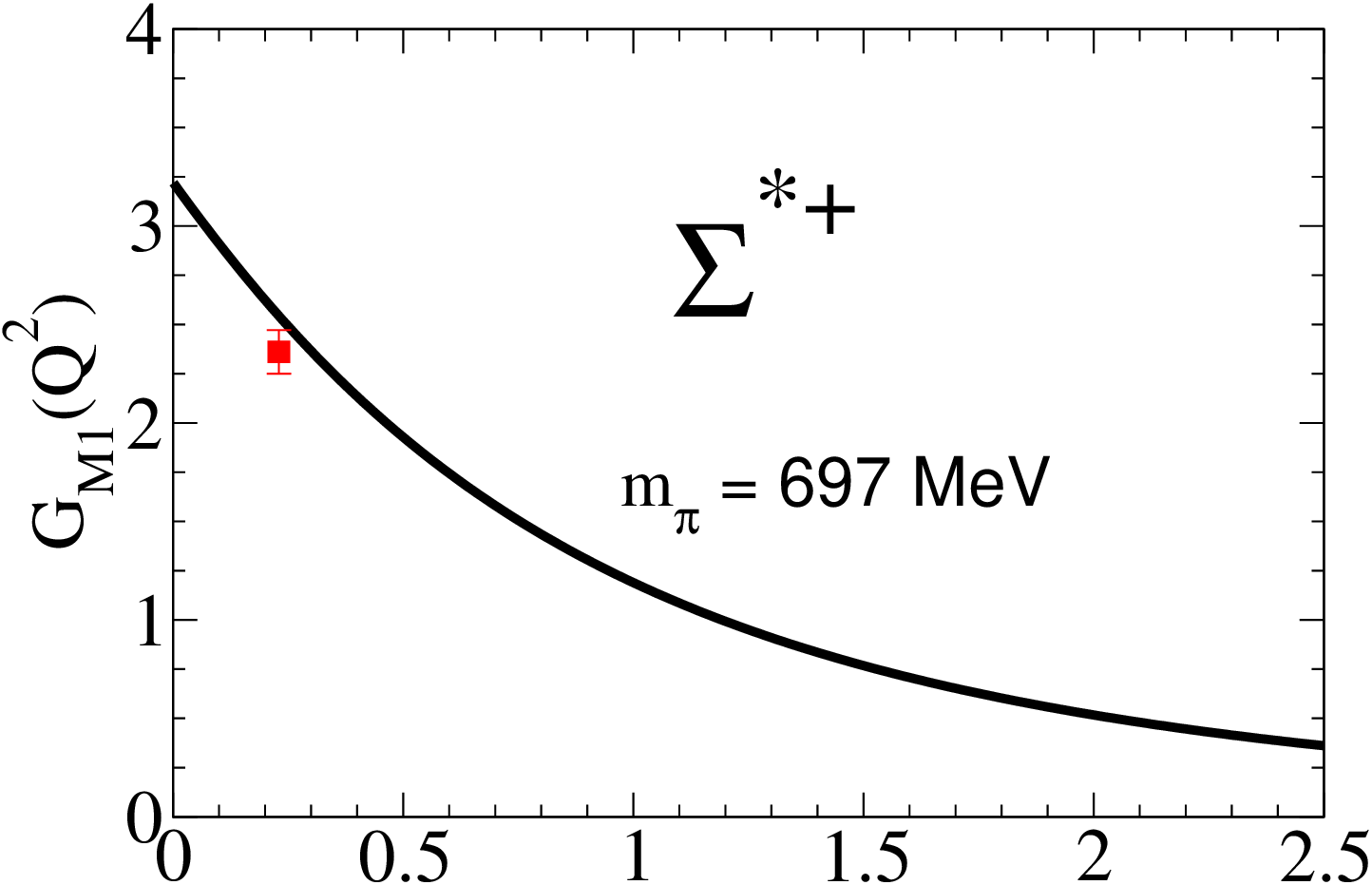} \hspace{.1cm}
\includegraphics[width=2.2in]{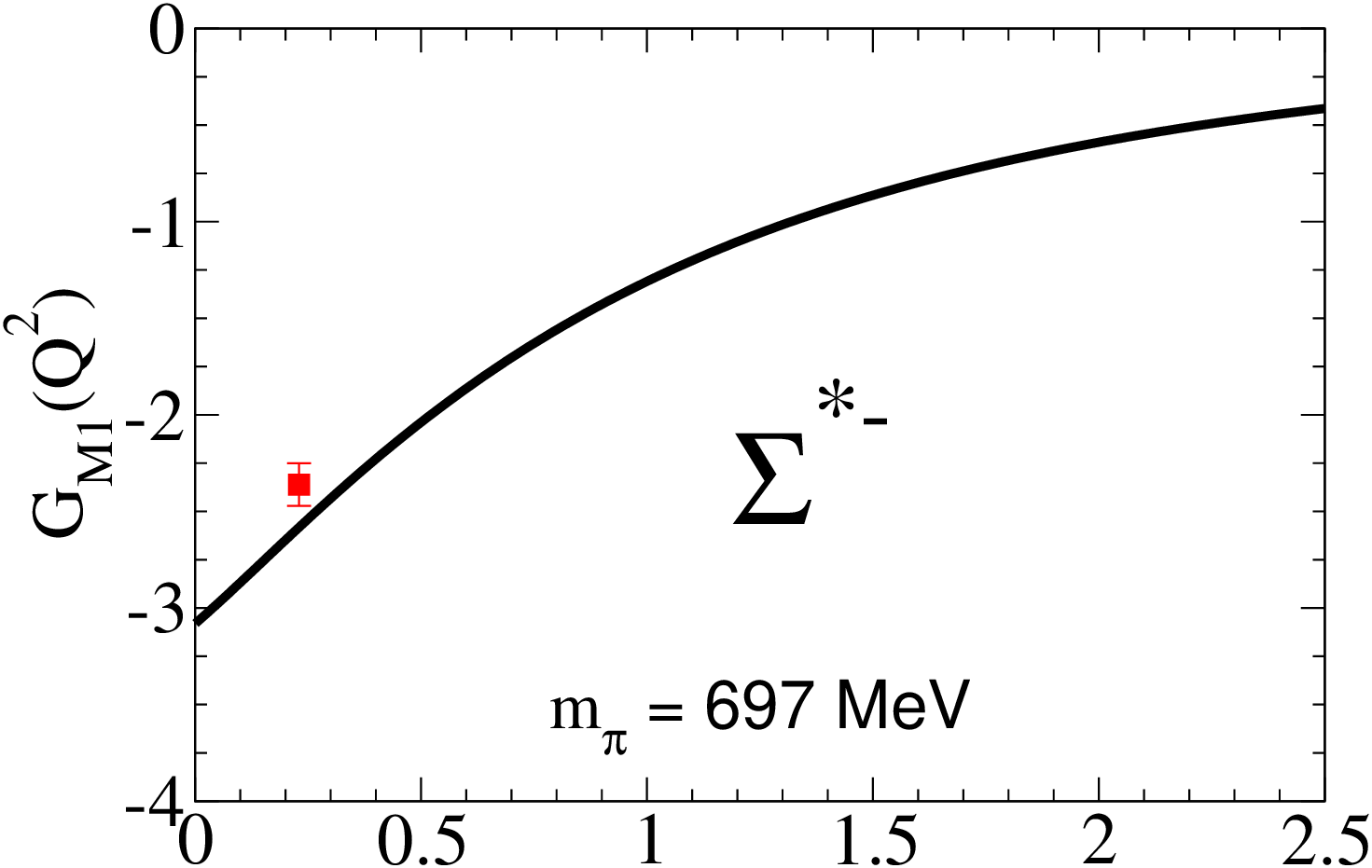} \hspace{.1cm}
\includegraphics[width=2.2in]{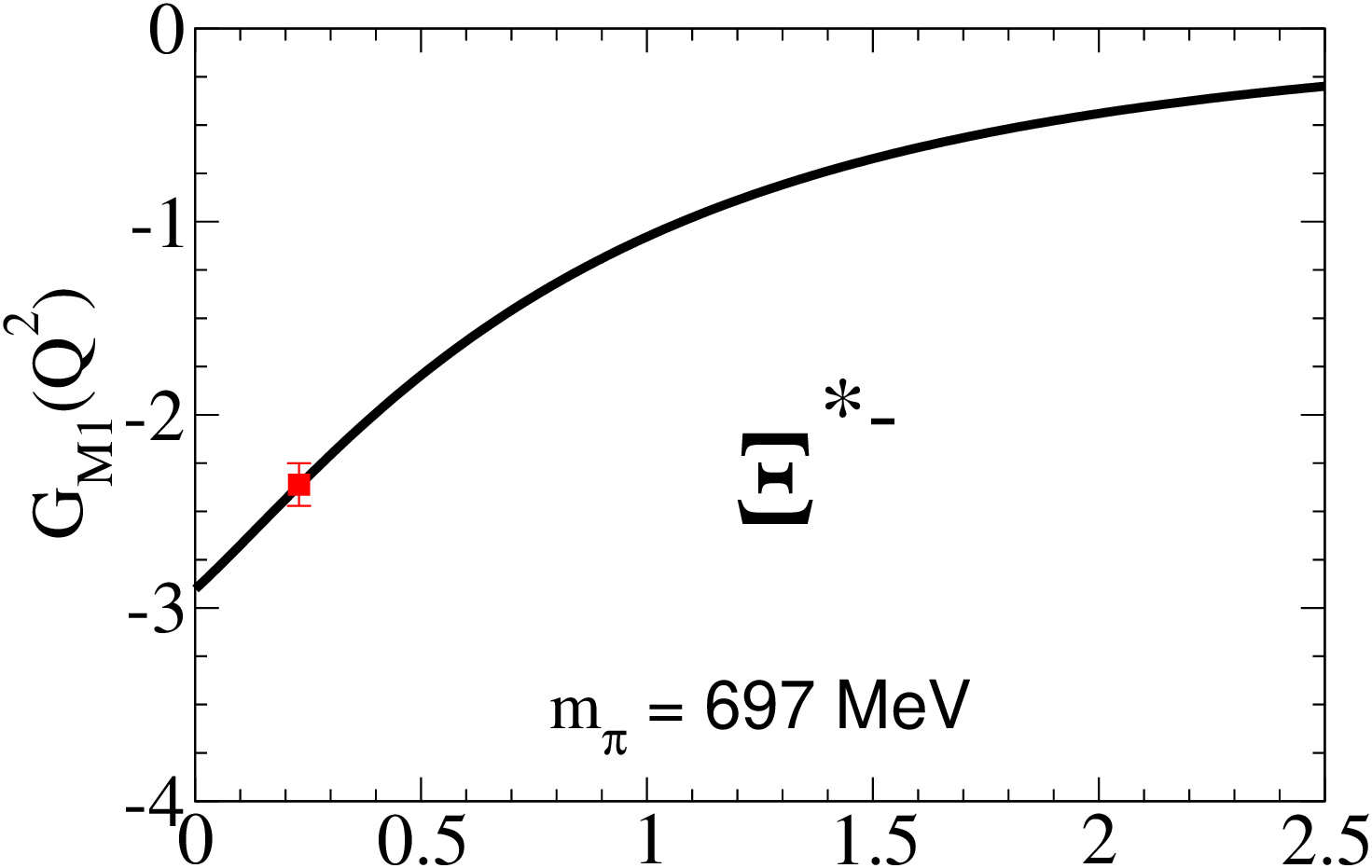}    \hspace{.1cm}
}}
\vspace{0.6cm}
\centerline{
\mbox{
\includegraphics[width=2.2in]{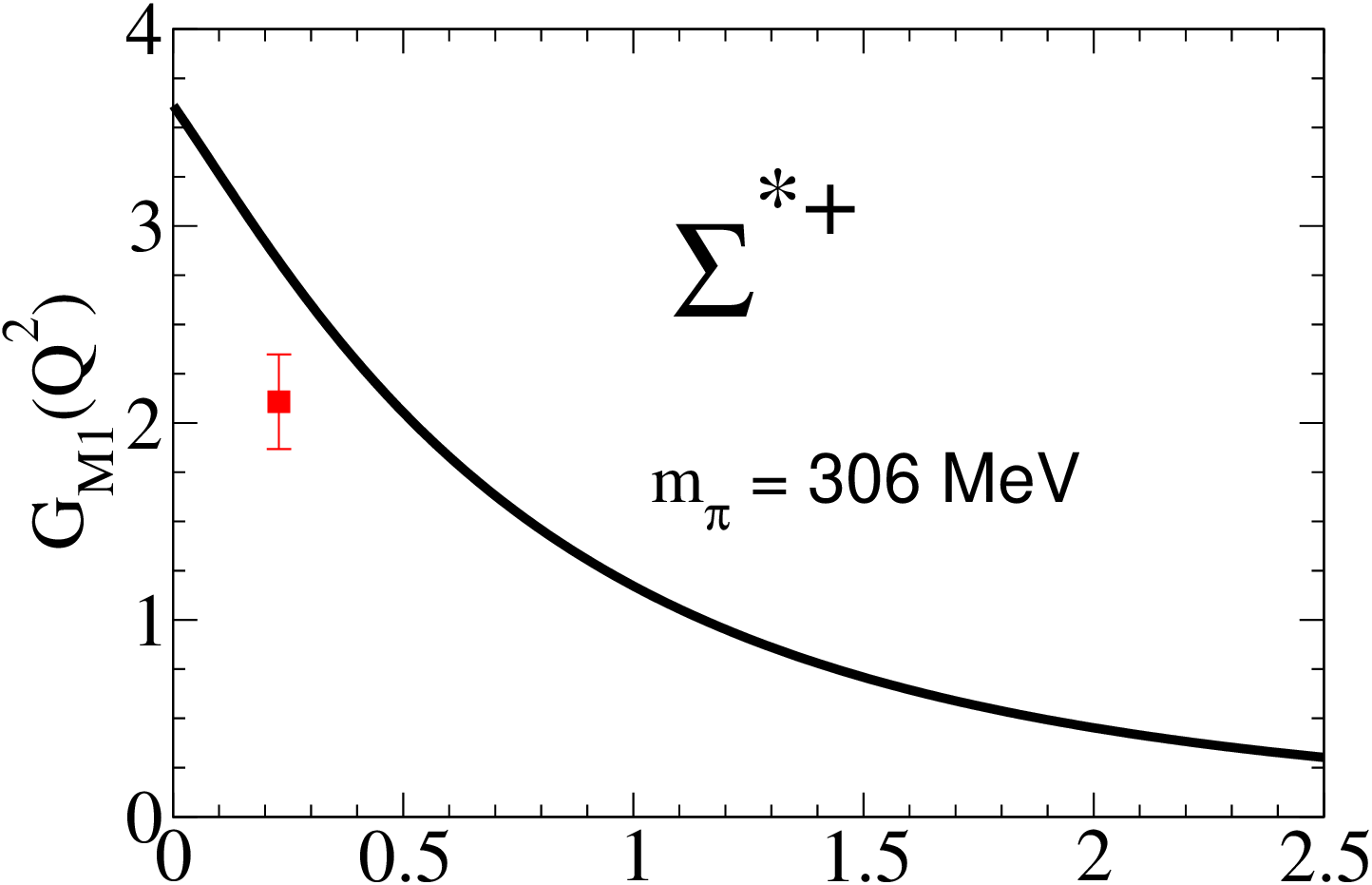} \hspace{.1cm}
\includegraphics[width=2.2in]{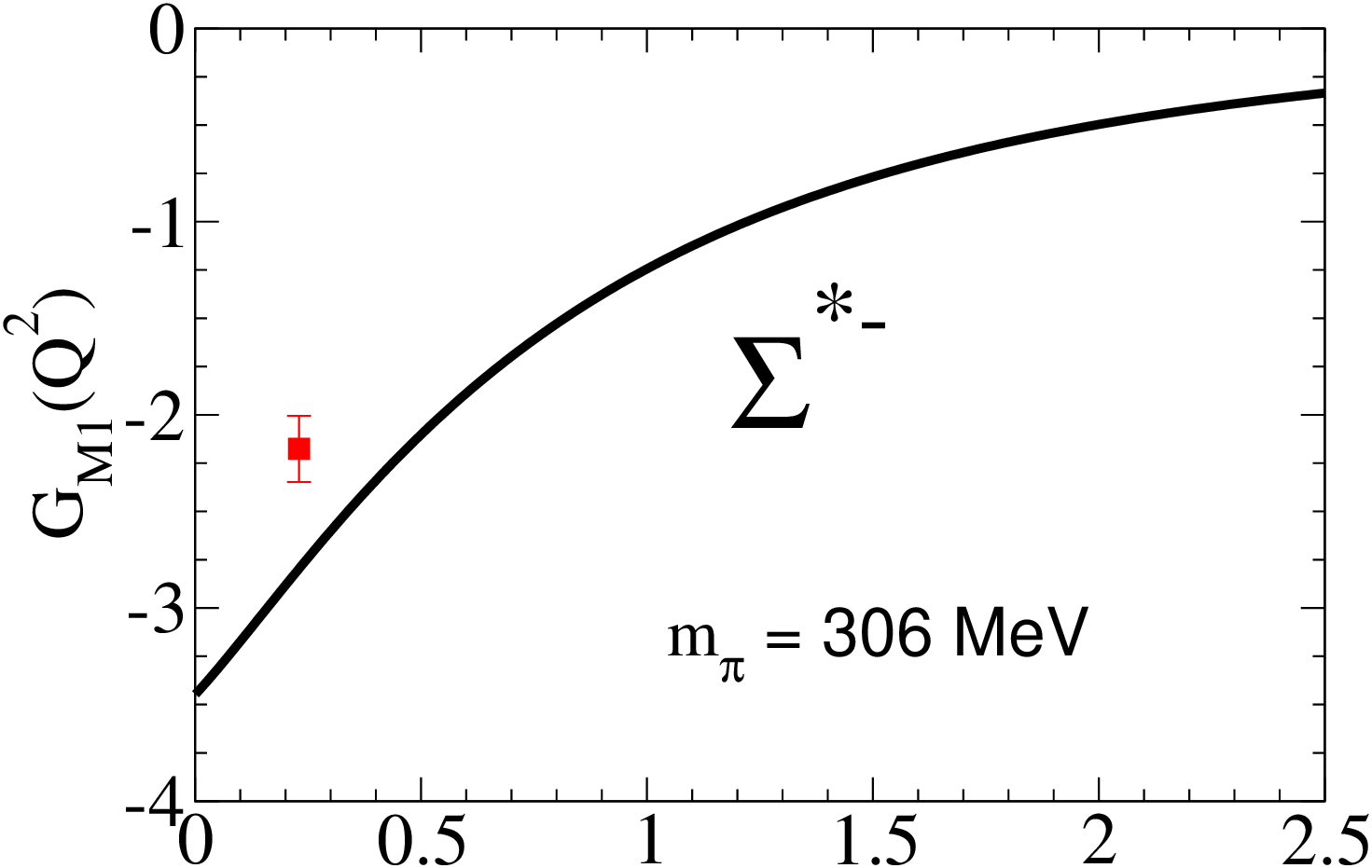} \hspace{.1cm}
\includegraphics[width=2.2in]{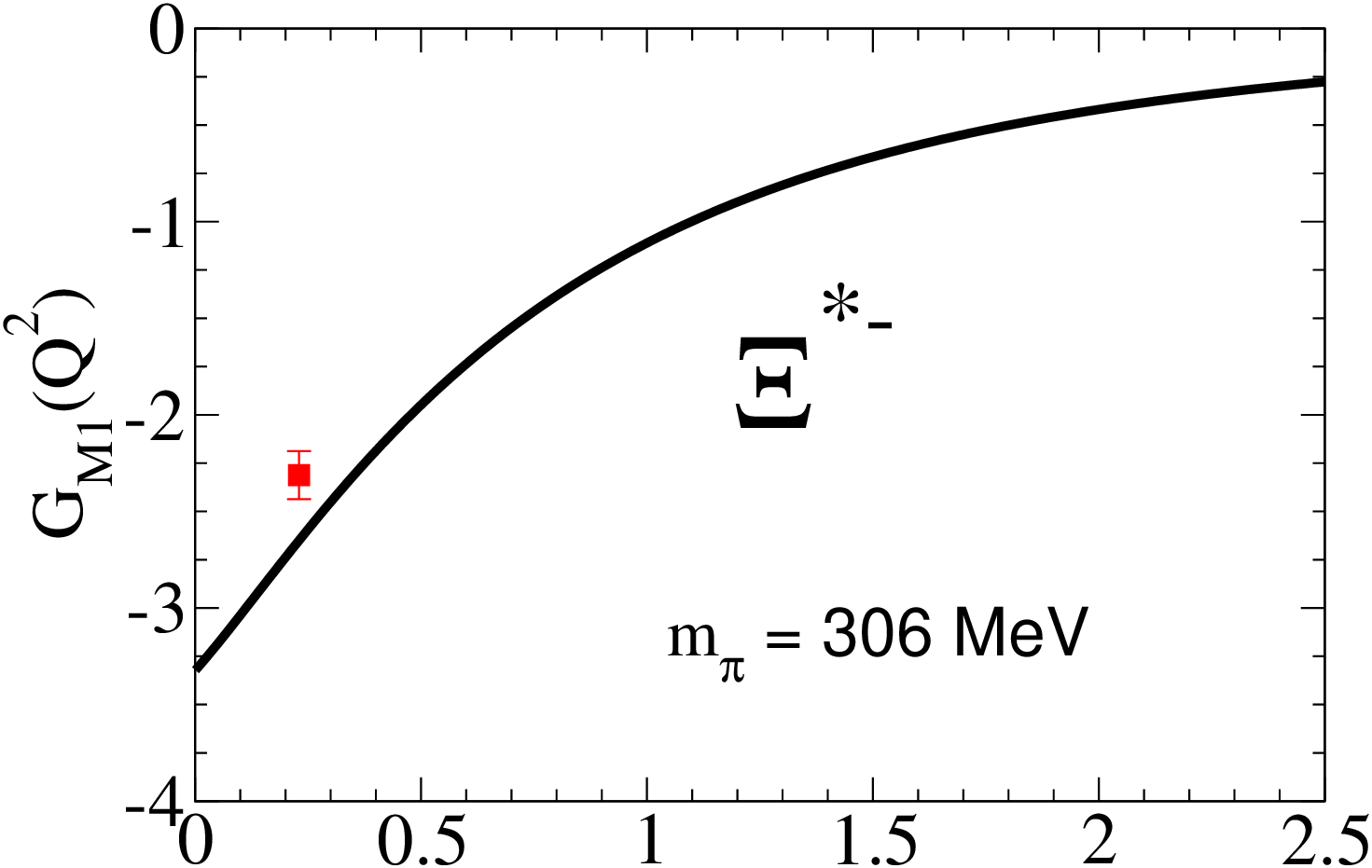}    \hspace{.1cm}
}}
\vspace{0.6cm}
\centerline{
\mbox{
\includegraphics[width=2.2in]{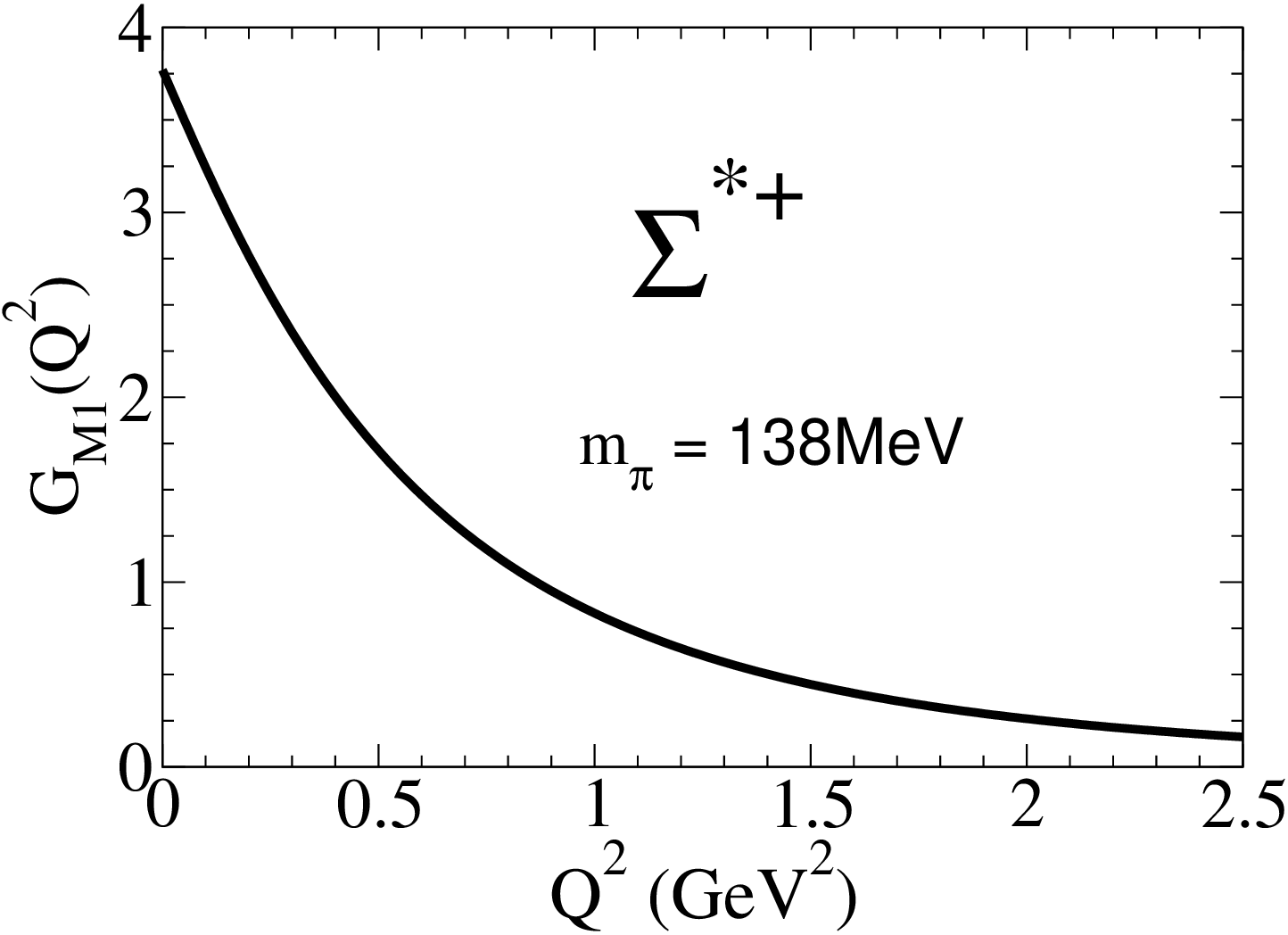}  \hspace{.1cm}
\includegraphics[width=2.2in]{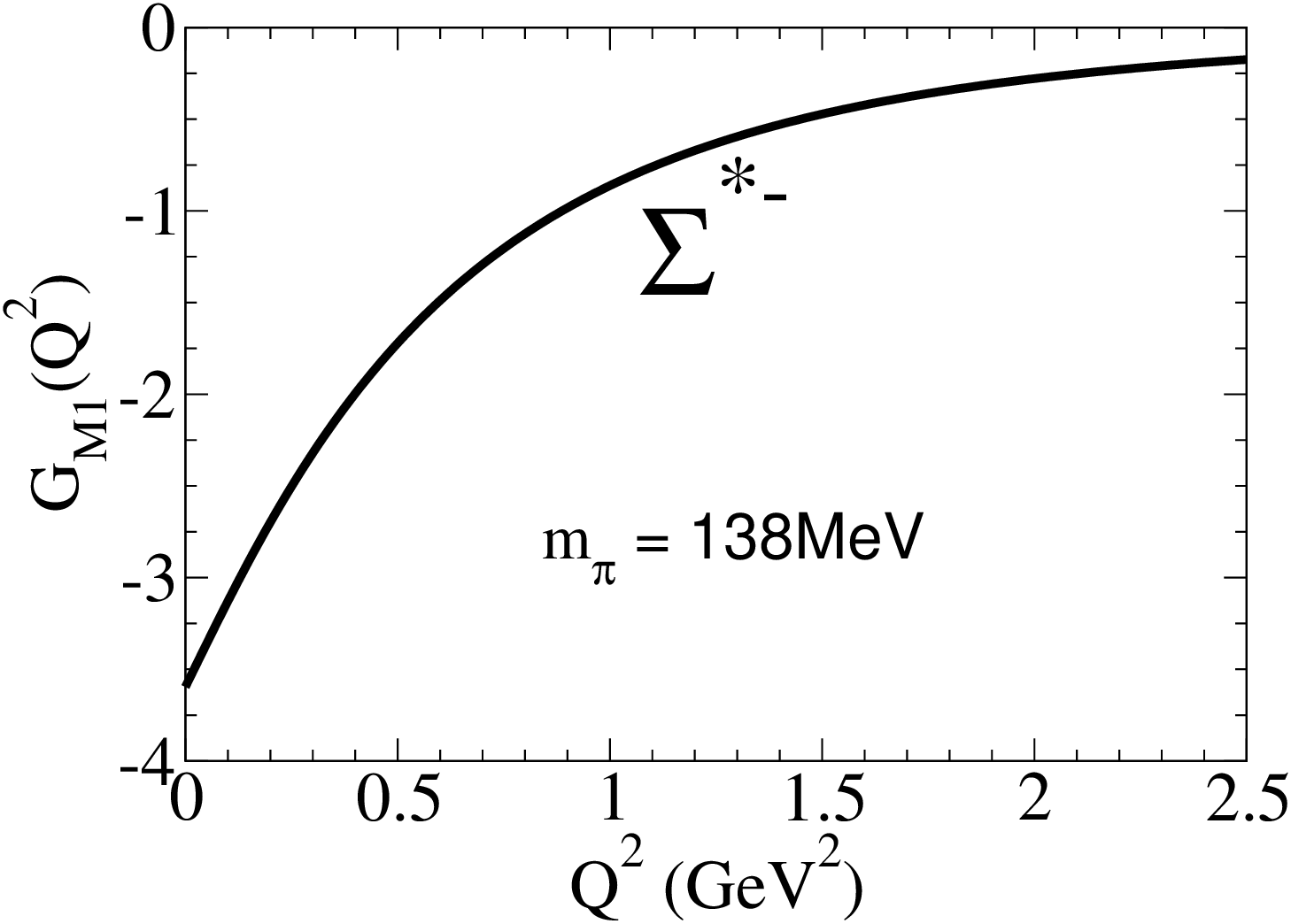}  \hspace{.1cm}
\includegraphics[width=2.2in]{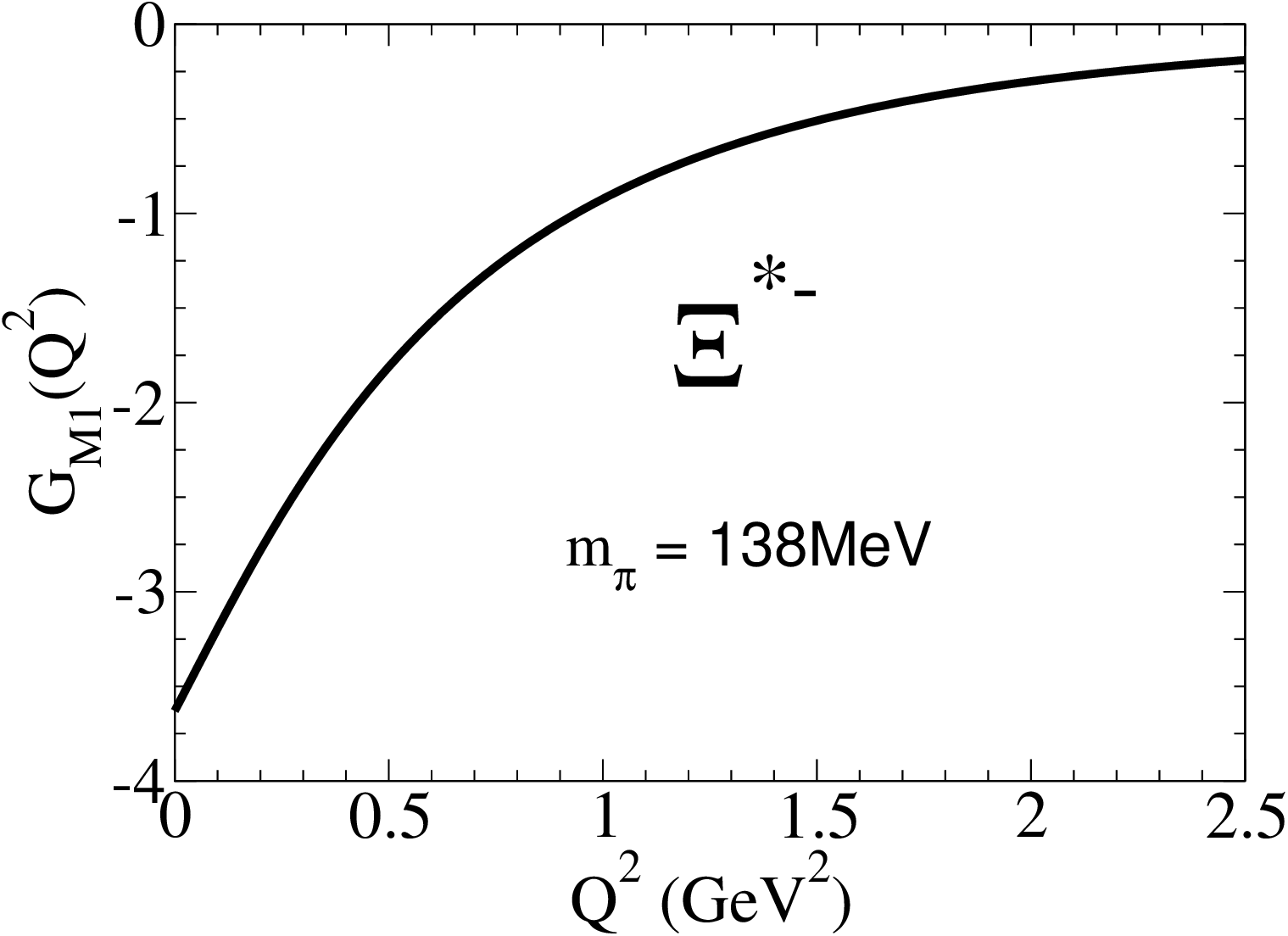}     \hspace{.1cm}
}}
\caption{\footnotesize{
$G_{E0}$ form factors for $\Sigma^{\ast +}$, $\Sigma^{\ast -}$
and     $\Xi^{\ast -}$. The quenched lattice QCD data is from
Ref.~\cite{Boinepalli09}.}}
\label{figGM1}
\end{figure*}

\section{Results}

To determine the parameters $c_0$, $d_0$
and $\alpha_2$ we 
have minimized
the $\chi^2$ for the
$G_{E0}$ and $G_{M1}$ form factor data
in quenched lattice QCD
from Ref.~\cite{Boinepalli09}.
The data are composed of 12 values of $m_\pi$
at one $Q^2$ ($Q^2=0.230$ GeV$^2$),
for $\Sigma^{\ast +}$, $\Sigma^{\ast -}$ and $\Xi^{\ast -}$.
The parameters $c_0$, $d_0$ and $\alpha_2$
are unconstrained, except for the condition
$\alpha_2 > 0$.
For $\kappa_s$ we keep the value $\kappa_s=1.462$
as determined by the physical $\Omega^-$ magnetic moment.
The results are presented in Fig.~\ref{figFit}.
The values obtained from the fit, together with the other parameters,
are presented in table~\ref{tableFit}.

The quality of the fit depends
on the baryon considered.
With the exclusion of the $\Xi^{\ast -}$
data for $G_{M1}$, the fit is excellent
for heavy pion masses; if we include $G_{M1}$ for the $\Xi^{*-}$  the fit is better for intermediate pion masses ($m_\pi=690-830$ MeV).
Because we are assuming that the momentum
range parameters are independent of $m_\pi$,
the failure of the model is expected
for some high pion mass.
As for the pion masses lower than 400 MeV,
we can expect more deviation, since the
effect of the pion cloud should become more important.
In particular, for the magnetic dipole form factors
it is known that the quenched data underestimate
the valence quark contribution, as well as the result
from full QCD (including meson cloud effects),
particularly for the case of the $\Delta$~\cite{Boinepalli09,Young}.
For the $\Sigma^\ast$ and $\Xi^\ast$, since the contribution
of the light quarks ($u$ or $d$) are smaller, the
effect of the pion cloud is also expected to be smaller.

\begin{table}[h,t]
\begin{center}
\begin{tabular}{l c  c}
\hline
\hline
$Q^2$ & $G_{E0}(Q^2)$ & $G_{M1}(Q^2)$ \\
\hline
    0.00 &      -1.000  &    -3.604 \\
    0.25 &      -0.752  &    -2.635 \\
    0.50 &      -0.544  &    -1.862 \\
    0.75 &      -0.393  &    -1.322 \\
    1.00 &      -0.287  &    -0.954 \\
    1.25 &      -0.213  &    -0.700 \\
    1.50 &      -0.160  &    -0.524 \\
    1.75 &      -0.122  &    -0.399 \\
    2.00 &      -0.094  &    -0.308 \\
    2.25 &      -0.074  &    -0.242 \\
    2.50 &      -0.058  &    -0.192 \\
\hline
\hline
\end{tabular}
\end{center}
\caption{Predictions for the $\Omega^-$ form factors.}
\label{tableOmega}
\end{table}

In Figs.~\ref{figGE0} and~\ref{figGM1}
we show the electric charge $G_{E0}$
and magnetic dipole $G_{M1}$ form factors
versus $Q^2$, for $\Sigma^{\ast +}$, $\Sigma^{\ast -}$ and $\Xi^{\ast -}$.
We present results for the following three cases: $m_\pi= 697$ MeV
corresponding to the SU(3) limit ($m_u=m_d=m_s$),
the lightest pion ($m_\pi= 306$ MeV),
and the physical point ($m_\pi=138$ MeV), which corresponds to our prediction.

Figure~\ref{figGE0} shows that a very good description is achieved for the
electric form factor even for $m_\pi=306$ MeV.
As for the magnetic dipole form factor shown in Fig.~\ref{figGM1},
the deviation of the model from the quenched lattice QCD data is noticeable,
particularly for the systems with two
light quarks, ($\Sigma^{\ast +}$ and $\Sigma^{\ast -}$).
One can see that the model results are closer
for $\Xi^{\ast -}$, particularly with
the value, $m_\pi= 697$ MeV.
We can interpret this deviation as
a consequence of fact that the meson cloud effect included in the quenched data has  the {\it wrong sign\/} (as has been observed for
$\Delta^+$~\cite{Boinepalli09,Young}).
For this reason it is natural to believe that
the quenched data will not only underestimate
the absolute value of the {\it exact\/}
magnetic form factors, but also the valence contributions
to  the magnetic form factors included in our calculation.

\begin{figure}[t]
\vspace{.5cm}
\centerline{
\mbox{
\includegraphics[width=3.1in]{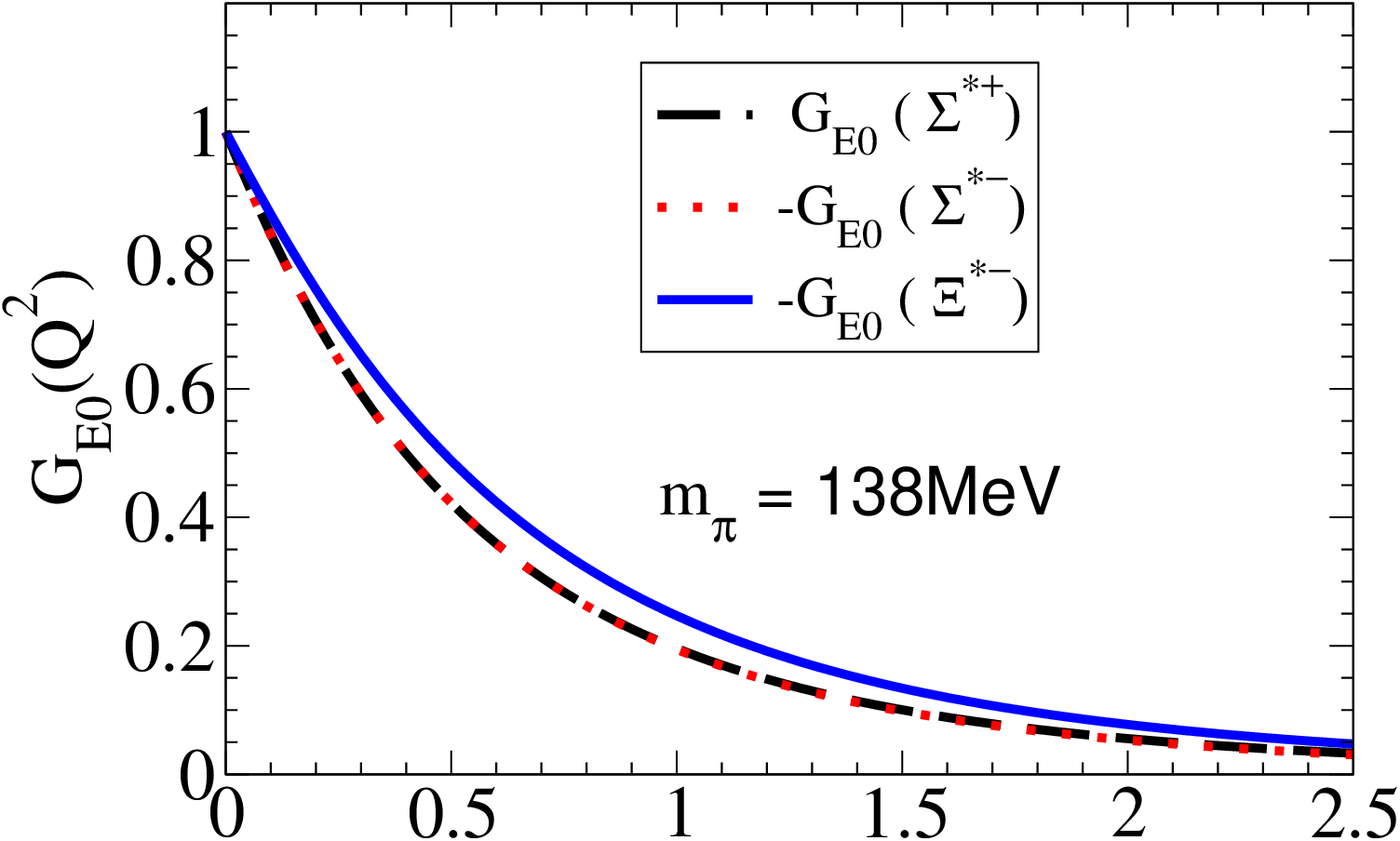}} \hspace{.1cm}}
\vspace{0.87cm}
\centerline{
\mbox{
\includegraphics[width=3.0in]{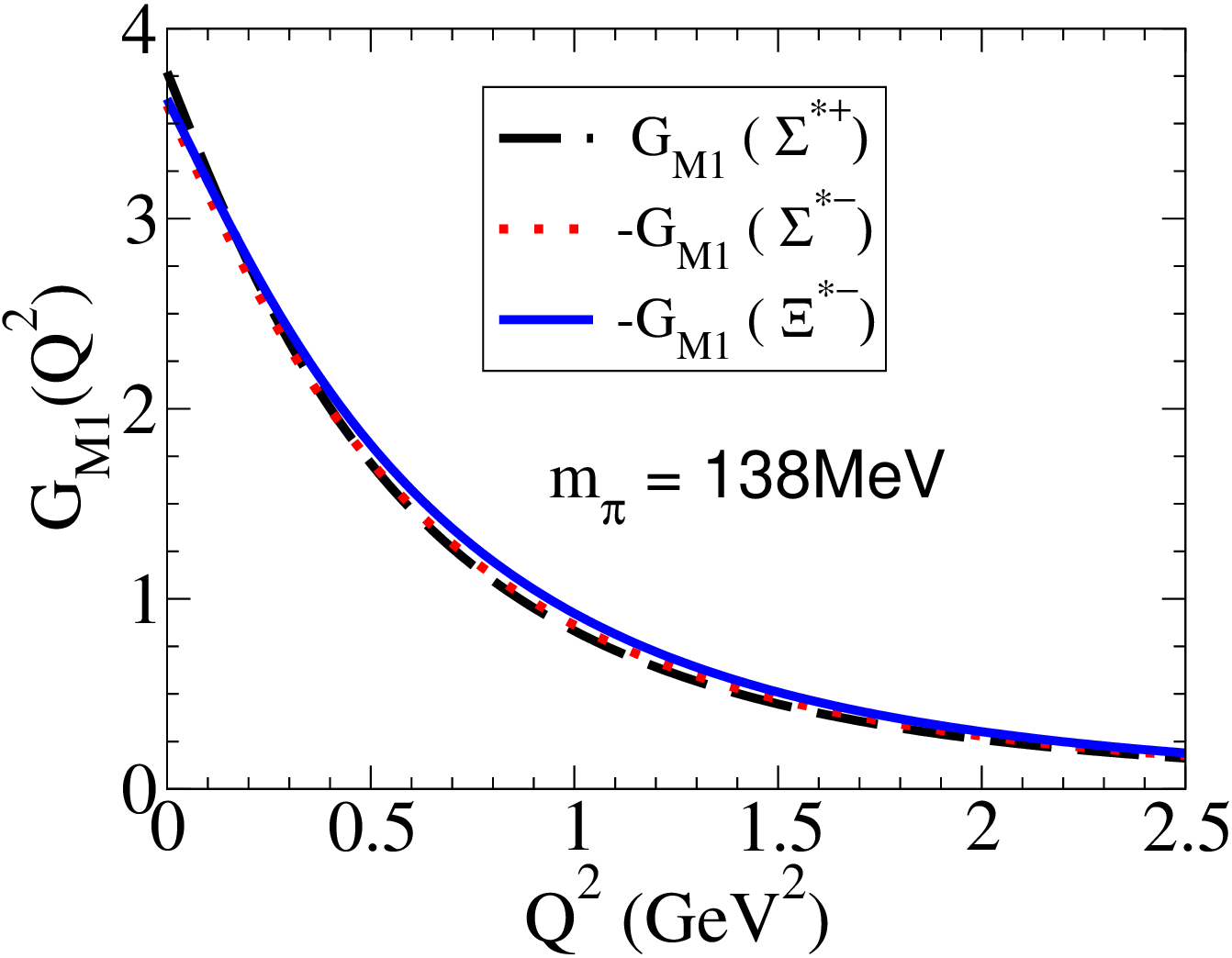}}}
\caption{\footnotesize{
Comparing $G_{\alpha}$ from $\Sigma^{\ast +}$
with $-G_{\alpha}$ from $\Sigma^{\ast -}$ and $\Xi^{\ast -}$.}}
\label{figGE0GM1}
\end{figure}

Figure~\ref{figGE0GM1} compares the form factors of the different baryons with each other.  We show our predictions for the absolute values $|G_{E0}|$ and $|G_{M1}|$
for $\Sigma^{\ast +}$
$\Sigma^{\ast -}$, and $\Xi^{\ast -}$ at the physical point ($m_\pi$=138 MeV).
The results for both form factors
suggest very similar
charge and magnetic moment distributions for the three baryons,
even though the parametrization
of the wave functions associated
with the strange quark are
substantially different
(compare  $\alpha_2$
with $\alpha_1$ in table \ref{tableFit}).
Another interesting point is that this similarity
implies that SU(3) symmetry
is approximately satisfied.
Only the differences in masses of the
baryons ($\Sigma^{\ast}$ and $\Xi^\ast$)
are responsible for the different values
of $|\mu_B|$.

\begin{figure}[t]
\vspace{.5cm}
\centerline{
\mbox{
\includegraphics[width=3.1in]{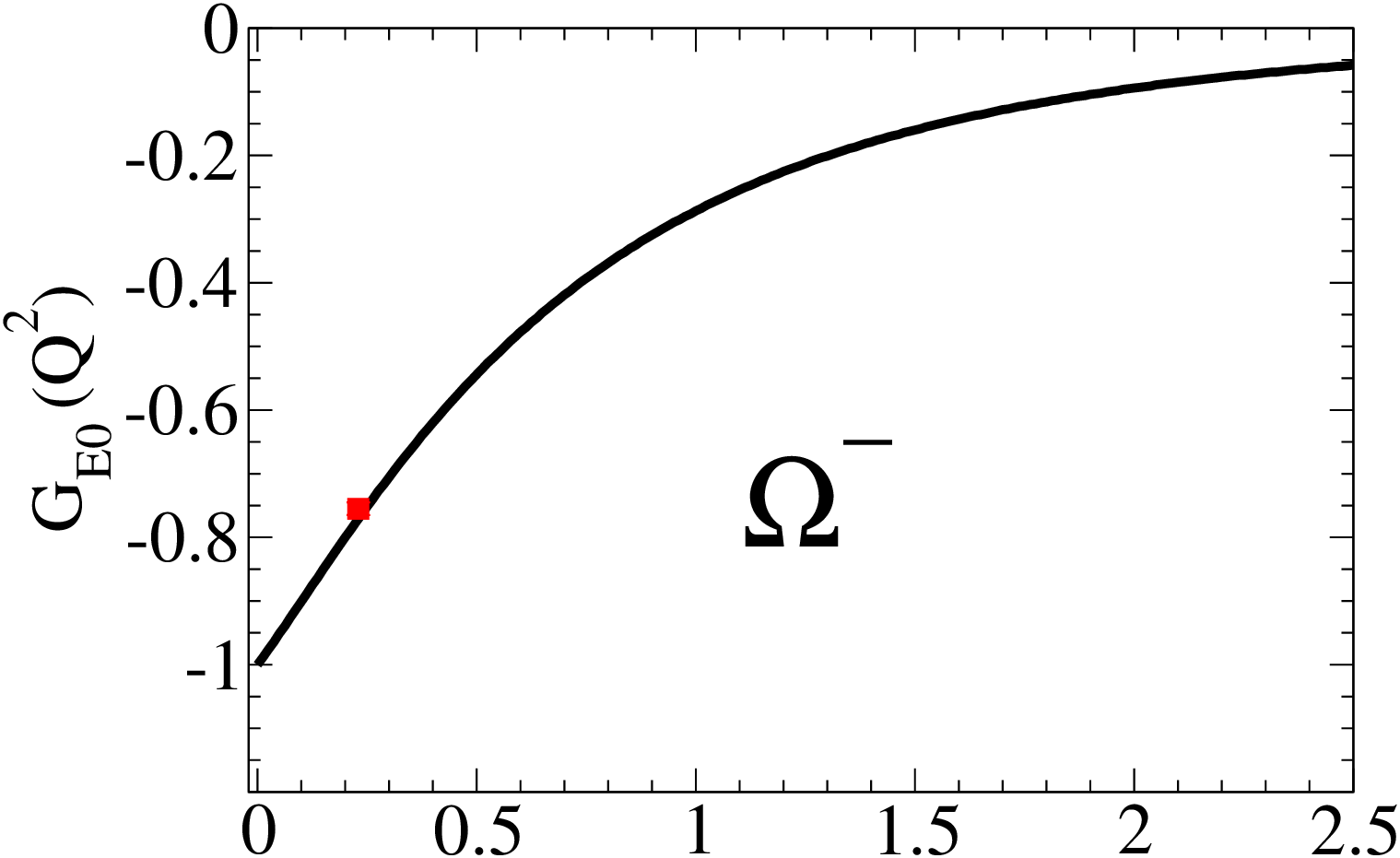}} \hspace{.1cm}}
\vspace{0.87cm}
\centerline{
\mbox{
\includegraphics[width=3.0in]{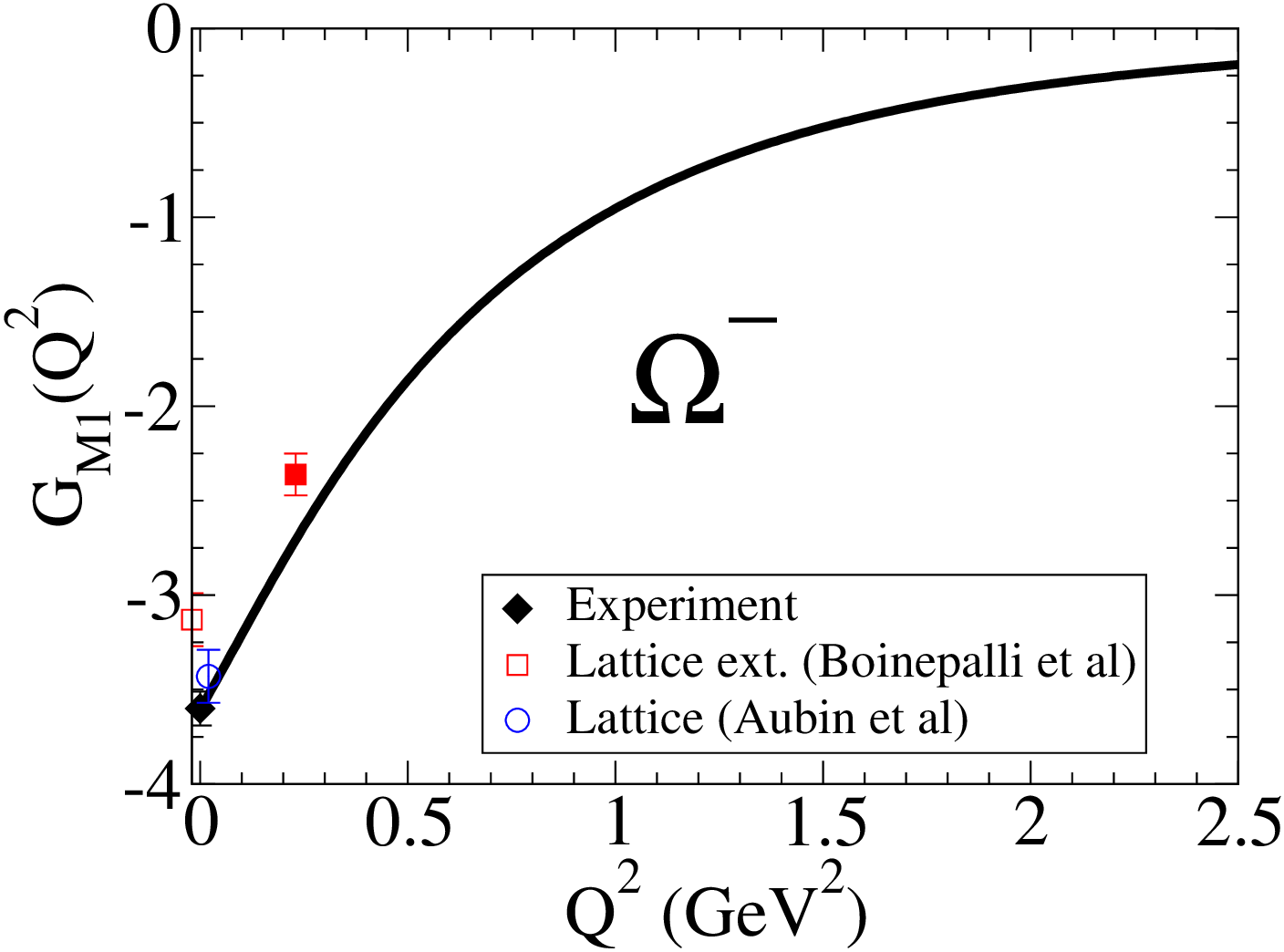}}}
\caption{\footnotesize{$\Omega^-$ form factors.
The squares represents the quenched
lattice data from Ref.~\cite{Boinepalli09}.
The circle represents the unquenched lattice data
from Ref.~\cite{Aubin}.  }}
\label{figOmega}
\end{figure}

Once the parameters
$c_0$, $d_0$ and $\alpha_2$ are adjusted to the lattice QCD data,
the $\Omega^-$ wave function at the physical $\Omega^-$ mass, and
the quark current with the physical quark masses
can be used to evaluate the $\Omega^-$ electromagnetic form factors.
The results are presented in the table
\ref{tableOmega} and in the Fig.~\ref{figOmega}.
At $Q^2=0$ the form factors are
constrained by
$e_{\Omega^-}=-1$ and by the experimental result for
$\mu_{\Omega^-}$, but
the evolution in $Q^2$ is a prediction.

Our results can be compared with the
results in Ref.~\cite{Boinepalli09} for
the $\Omega^-$ magnetic dipole form factor
at $Q^2=0.230$ GeV$^2$, extracted from
the simulation for the $\Delta^-$ at the
pion mass $m_\pi =697$ MeV:
$G_{M1}(Q^2)=-2.36\pm 0.11$.
The experimental value (used in our calibration)  which gives
$G_{M1}(0)= \mu_{\Omega^-}\sfrac{M_\Omega}{M_N}=-3.60 \pm 0.10$
with $\mu_{\Omega^-}$ in units $\mu_N$ \cite{Wallace95},
is also presented in the graph.
In addition, there are unquenched lattice data at $Q^2=0$ from
Aubin \etal~\cite{Aubin}:
$G_{M1}(0)= -3.44\pm0.14$, and the
extrapolation from the Ref.~\cite{Boinepalli09}
for $Q^2=0$: $G_{M1}(0)= -3.14\pm0.12$.

Finally, our model 
predicts the $\Omega^-$
squared radii of
\mbox{$<r_{E0}^2>= 0.22$ fm$^2$} and
$<r_{M1}^2>= 0.27$ fm$^2$.
These values are close to
those estimated by the lattice QCD data in Ref.~\cite{Boinepalli09},
which we have used
to calibrate the $s$ quark current and momentum distribution.
In that work the corresponding results are:
$<r_{E0}^2> = <r_{M1}^2> = 0.307\pm 0.015$ fm$^2$.
It is expected that inclusion of the 
meson cloud will increase that value \cite{Boinepalli09}.
Other estimates of the $\Omega^-$ charge radius
\cite{Kunz89,Schwesinger92,Barik95,Wagner00,Sahoo95,Buchmann07,Gobbi92}
lead to a result between 0.16 fm$^2$ and 0.61 fm$^2$,
but not all the estimates are consistent
with the experimental $\Omega^-$ magnetic moment.
A chiral quark model with consistent exchange currents \cite{Wagner00}  that agrees with the data to
a precision of better than 6\% predicts $<r_{E0}^2>=0.61$ fm$^2$.

\section{Conclusions}

In this work we have extended the covariant
spectator quark model (based on the Covariant Spectator Theory) to include the strange quark.  
We have chosen to study the baryon decuplet as a first application of the model because the structure (symmetric spin 3/2 state) is
simpler than the baryon octet, where the spin 1/2 structure requires a mixture of diquark states with
spin 0 and 1~\cite{Nucleon}.
Using the measured $\Omega^-$ magnetic moment and the recent lattice QCD data for the baryon decuplet~\cite{Boinepalli09} we have fixed the new model parameters associated with the $s$ quark contributions to the  current and the baryon wave functions.

The experimental result for $\mu_{\Omega^-}$
fixes the anomalous magnetic moment of the strange quark, $\kappa_s$,
at a value different from
$\sfrac{1}{2}(\kappa_u+\kappa_d)$,
suggesting a violation of the SU(3) symmetry
at the level of 20\%.

In this first study we have restricted the model
to the description of the dominant form factors: $G_{E0}$ and $G_{M1}$.
The subleading form factors $G_{E2}$ and $G_{M3}$
are also interesting for future study.
For example, the subleading form factors can emerge in the present model
when the D-states are included in the spin 3/2 systems.
Although the D-states are important in the electromagnetic
transition between the octet and the decuplet baryons
(as an example for the $\gamma N \to \Delta$ transition),
the D-states are not expected to be dominant
for $G_{E0}$ and $G_{M1}$,
as was shown for the $\Delta$ case~\cite{DeltaFF,DeltaDFF,DeltaDFF2}.

After calibration of the model, we have predicted
the $\Omega^-$ form factors. The $\Omega^-$ is a
very interesting object to study since it is
composed of three strange quarks.
Lattice QCD simulation for the $\Omega^-$
form factors can presently be performed at the
physical strange quark mass~\cite{Boinepalli09,Aubin}.
As the $\Omega^-$ is considerably more stable than
the octet baryons and other members of the decuplet baryons
except for the nucleons, there is a hope that
the  $\Omega^-$ form factors
can be measured in the 
near future.

Our work so far is restricted to
the valence quark degrees of freedom.
This framework can be regarded as a good approximation
since the meson cloud effects,
the pion could as well as those of kaons,
are expected to be smaller,
when strange quarks are present in the baryon.
This statement is more valid for
higher $Q^2$ and larger pion masses.
The $\Omega^-$
meson cloud is expected to be
small, although the difference
between the extrapolation with
the quenched lattice QCD data from
Ref.~\cite{Boinepalli09} and the experimental result
indicates that the meson cloud is not negligible.
A possible reason for this is that the quenched calculations
do not include the virtual transitions $\Omega^- \to \Xi K$
due to the omission of light quark loops (only the $s$ quark is considered).
%
%
Finally the discrepancy can also be partially
due to the $\Omega^-$ mass, that in
the quenched lattice simulation exceeds
the physical mass by 3.6\%.
In future we plan to study these meson cloud effects.


We can also extend the present model to the study of the octet baryons, and
to the heavy quark sector ($c$ and $b$ quarks), where valence quark degrees of
freedom dominate.

\vspace{0.2cm}
\noindent
{\bf Acknowledgments:}

G.~R. thanks David Richard, Huey-Wen Lin, Christopher Thomas
and Nilmani Mathur for helpful discussions.
This work was partially support by Jefferson Science Associates,
LLC under U.S. DOE Contract No. DE-AC05-06OR23177.
G.~R.\ was supported by the Portuguese Funda\c{c}\~ao para
a Ci\^encia e Tecnologia (FCT) under Grant No.\
SFRH/BPD/26886/2006.
This work has been supported in part by the European Union
(HadronPhysics2 project ``Study of strongly interacting matter'').

\begin{thebibliography}{00}





\bibitem{Yao06}
  W.~M.~Yao {\it et al.}  [Particle Data Group],
  J.\ Phys.\ G {\bf 33}, 1 (2006). 



\bibitem{Beg64}
  M.~A.~B.~Beg, B.~W.~Lee and A.~Pais,
  Phys.\ Rev.\ Lett.\  {\bf 13}, 514 (1964).

\bibitem{Hikasa92}
  K.~Hikasa {\it et al.}  [Particle Data Group],
  Phys.\ Rev.\  D {\bf 45}, S1, (1992) (see p.~482);
  [Erratum-ibid.\  D {\bf 46}, 5210 (1992)].


\bibitem{Muller}
   Greiner Muller,
   {\bf Quantum Mechanics Symmetries},
   Spring-Verlag Berlin Heidelberg (1994).




\bibitem{Das80}
  T.~Das and S.~P.~Misra,
  Phys.\ Lett.\  B {\bf 96}, 165 (1980).

\bibitem{Tomozawa82}
  Y.~Tomozawa,
  Phys.\ Rev.\  D {\bf 25}, 795 (1982).

\bibitem{Georgi83}
  H.~Georgi and A.~Manohar,
  Phys.\ Lett.\  B {\bf 132}, 183 (1983).


\bibitem{Verma87}
  R.~C.~Verma and M.~P.~Khanna,
  Phys.\ Lett.\  B {\bf 183}, 207 (1987).

\bibitem{Krivoruchenko87}
  M.~I.~Krivoruchenko,
  Sov.\ J.\ Nucl.\ Phys.\  {\bf 45}, 109 (1987)
  [Yad.\ Fiz.\  {\bf 45}, 169 (1987)].


\bibitem{Kim89}
  J.~H.~Kim, C.~H.~Lee and H.~K.~Lee,
  Nucl.\ Phys.\  A {\bf 501} (1989) 835.


\bibitem{Kunz89}
  J.~Kunz and P.~J.~Mulders,
  Phys.\ Lett.\  B {\bf 231}, 335 (1989);
  J.~Kunz and P.~J.~Mulders,
  Phys.\ Rev.\  D {\bf 41}, 1578 (1990).

\bibitem{Chao90}
  K.~T.~Chao,
  Phys.\ Rev.\  D {\bf 41}, 920 (1990).




\bibitem{Deihl91}
  H.~T.~Deihl {\it et al.}
  Phys.\ Rev.\ Lett.\  {\bf 67}, 804 (1991).


\bibitem{Wallace95}
  N.~B.~Wallace {\it et al.},
  Phys.\ Rev.\ Lett.\  {\bf 74}, 3732 (1995).




\bibitem{Schwesinger92}
  B.~Schwesinger and H.~Weigel,
  Nucl.\ Phys.\  A {\bf 540}, 461 (1992).


\bibitem{Hong94}
  S.~T.~Hong and G.~E.~Brown,
  Nucl.\ Phys.\  A {\bf 580}, 408 (1994).


\bibitem{Ha98}
  P.~Ha,
  Phys.\ Rev.\  D {\bf 58}, 113003 (1998)
  [arXiv:hep-ph/9804383].



\bibitem{Linde98}
  J.~Linde, T.~Ohlsson and H.~Snellman,
  Phys.\ Rev.\  D {\bf 57}, 5916 (1998)
  [arXiv:hep-ph/9709468].

\bibitem{Zhu98}
  S.~L.~Zhu, W.~Y.~P.~Hwang and Z.~S.~P.~Yang,
  Phys.\ Rev.\  D {\bf 57}, 1527 (1998)
  [arXiv:hep-ph/9802322].



\bibitem{Schlumpf93}
  F.~Schlumpf,
  Phys.\ Rev.\  D {\bf 48}, 4478 (1993)
  [arXiv:hep-ph/9305293].


\bibitem{Barik95}
  N.~Barik, P.~Das and A.~R.~Panda,
  Pramana {\bf 44} (1995) 145.


\bibitem{Wagner00}
  G.~Wagner, A.~J.~Buchmann and A.~Faessler,
  J.\ Phys.\ G {\bf 26}, 267 (2000).


\bibitem{Aliev00}
  T.~M.~Aliev, A.~Ozpineci and M.~Savci,
  Nucl.\ Phys.\  A {\bf 678}, 443 (2000)
  [arXiv:hep-ph/0002228].



\bibitem{Iqubal00}
  A.~Iqubal, M.~Dey and J.~Dey
  Phys.\ Lett.\  B {\bf 477}, 125 (2000)
  [arXiv:hep-ph/9906479].


\bibitem{Kerbikov00}
  B.~O.~Kerbikov and Yu.~A.~Simonov,
  Phys.\ Rev.\  D {\bf 62}, 093016 (2000)
  [arXiv:hep-ph/0001243].


\bibitem{Franklin02}
  J.~Franklin,
  Phys.\ Rev.\  D {\bf 66}, 033010 (2002).



\bibitem{Sahu02}
  S.~Sahu,
  Rev.\ Mex.\ Fis.\  {\bf 48}, 48 (2002).

\bibitem{An06}
  C.~S.~An, Q.~B.~Li, D.~O.~Riska and B.~S.~Zou,
  Phys.\ Rev.\  C {\bf 74}, 055205 (2006)
  [Erratum-ibid.\  C {\bf 75}, 069901 (2007)]
  [arXiv:nucl-th/0610009].


\bibitem{Ledwig08}
  T.~Ledwig, A.~Silva and M.~Vanderhaeghen,
  Phys.~Rev.~D {\bf 79}, 094025 (2009)
  arXiv:0811.3086 [hep-ph].


\bibitem{Aliev09}
  T.~M.~Aliev, K.~Azizi and M.~Savci,
  arXiv:0904.2485 [hep-ph].


\bibitem{Leonard90}
  W.~J.~Leonard and W.~J.~Gerace,
  Phys.\ Rev.\  D {\bf 41}, 924 (1990).




\bibitem{Gershtein81}
  S.~S.~Gershtein and Yu.~M.~Zinovev,
  Sov.\ J.\ Nucl.\ Phys.\  {\bf 33}, 772 (1981)
  [Yad.\ Fiz.\  {\bf 33}, 1442 (1981)].

\bibitem{Richard82}
  J.~M.~Richard,
  Z.\ Phys.\  C {\bf 12}, 369 (1982).


\bibitem{Isgur82}
  N.~Isgur, G.~Karl and R.~Koniuk,
  Phys.\ Rev.\  D {\bf 25}, 2394 (1982).


\bibitem{Krivoruchenko91}
  M.~I.~Krivoruchenko and M.~M.~Giannini,
  Phys.\ Rev.\  D {\bf 43}, 3763 (1991).


\bibitem{Butler94}
  M.~N.~Butler, M.~J.~Savage and R.~P.~Springer,
  Phys.\ Rev.\  D {\bf 49}, 3459 (1994)
  [arXiv:hep-ph/9308317].

\bibitem{Sahoo95}
  R.~K.~Sahoo, A.~R.~Panda and A.~Nath,
  Phys.\ Rev.\  D {\bf 52}, 4099 (1995).


\bibitem{Buchmann02}
   A.~J.~Buchmann and E.~M.~Henley,
   Phys.\ Rev.\  D {\bf 65}, 073017 (2002);
  A.~J.~Buchmann and R.~F.~Lebed,
  Phys.\ Rev.\  D {\bf 67}, 016002 (2003)
  [arXiv:hep-ph/0207358].


\bibitem{Buchmann07}
  A.~J.~Buchmann,
  arXiv:0712.4383 [hep-ph].

\bibitem{Geng09}
  L.~S.~Geng, J.~Martin Camalich and M.~J.~Vicente Vacas,
  Phys.\ Rev.\  D {\bf 80}, 034027 (2009)
  [arXiv:0907.0631 [hep-ph]].


\bibitem{Gobbi92}
  C.~Gobbi, S.~Boffi and D.~O.~Riska,
  Nucl.\ Phys.\  A {\bf 547}, 633 (1992).







\bibitem{Buchmann08}
  A.~J.~Buchmann and E.~M.~Henley,
  Eur.\ Phys.\ J.\  A {\bf 35}, 267 (2008)
  [arXiv:0808.1165 [hep-ph]].











\bibitem{Bernard82}
  C.~W.~Bernard, T.~Draper, K.~Olynyk and M.~Rushton,
  Phys.\ Rev.\ Lett.\  {\bf 49}, 1076 (1982).






\bibitem{Leinweber92}
  D.~B.~Leinweber, T.~Draper and R.~M.~Woloshyn,
  Phys.\ Rev.\  D {\bf 46}, 3067 (1992)
  [arXiv:hep-lat/9208025].




\bibitem{Lee05}
  F.~X.~Lee, R.~Kelly, L.~Zhou and W.~Wilcox,
  Phys.\ Lett.\  B {\bf 627}, 71 (2005)
  [arXiv:hep-lat/0509067].



\bibitem{Aubin}
  C.~Aubin, K.~Orginos, V.~Pascalutsa and M.~Vanderhaeghen,
  Phys.\ Rev.\ D {\bf 79}, 051502(R) (2009)
  arXiv:0811.2440 [hep-lat].



\bibitem{Boinepalli09}
  S.~Boinepalli, D.~B.~Leinweber, P.~J.~Moran,
  A.~G.~Williams, J.~M.~Zanotti and J.~B.~Zhang,
  arXiv:0902.4046 [hep-lat].





\bibitem{Lin08}
  H.~W.~Lin {\it et al.}  [Hadron Spectrum Collaboration],
  Phys.\ Rev.\  D {\bf 79}, 034502 (2009)
  [arXiv:0810.3588 [hep-lat]].


\bibitem{Tiburzi08}
  B.~C.~Tiburzi and A.~Walker-Loud,
  Phys.\ Lett.\  B {\bf 669}, 246 (2008).
  [arXiv:0808.0482 [nucl-th]].


\bibitem{Aoki09}
  S.~Aoki {\it et al.}  [PACS-CS Collaboration and PACS-CS
                  Collaboration and PACS-CS Collaboration],
  Phys.\ Rev.\  D {\bf 79}, 034503 (2009)
  arXiv:0905.0962 [hep-lat].


\bibitem{Drach09}
  V.~Drach {\it et al.},
  PoS {\bf LATTICE2008}, 123 (2008)
  [arXiv:0905.2894 [hep-lat]].


\bibitem{Gross}
  F.~Gross,
  Phys.\ Rev.\  {\bf 186}, 1448 (1969);
  F.~Gross, J.~W.~Van Orden and K.~Holinde,
  Phys.\ Rev.\ C {\bf 45}, 2094 (1992).


\bibitem{FixedAxis}
  F.~Gross, G.~Ramalho and M.~T.~Pe\~na,
  Phys.\ Rev.\  C {\bf 77}, 035203 (2008).




\bibitem{NDelta}
  G.~Ramalho, M.~T.~Pe\~na and F.~Gross,
  Eur.\ Phys.\ J.\  A {\bf 36}, 329 (2008)
  [arXiv:0803.3034 [hep-ph]].


\bibitem{NDeltaD}
  G.~Ramalho, M.~T.~Pe\~na and F.~Gross,
  Phys.\ Rev.\  D {\bf 78}, 114017 (2008)
  [arXiv:0810.4126 [hep-ph]].



\bibitem{LatticeD}
  G.~Ramalho and M.~T.~Pena,
  Phys.\ Rev.\  D {\bf 80}, 013008 (2009)
  [arXiv:0901.4310 [hep-ph]].



\bibitem{Nucleon}
  F.~Gross, G.~Ramalho and M.~T.~Pe\~na,
  Phys.\ Rev.\  C {\bf 77}, 015202 (2008)
  [arXiv:nucl-th/0606029].



\bibitem{DeltaFF}
  G.~Ramalho and M.~T.~Pe\~na,
  J.~Phys.~G {\bf 56}, 0805004 (2009)
  [arXiv:0807.2922 [hep-ph]].


\bibitem{DeltaDFF}
  G.~Ramalho, M.~T.~Pena and F.~Gross,
  Phys.\ Lett.\  B {\bf 678}, 355 (2009)
  [arXiv:0902.4212 [hep-ph]].



\bibitem{Gross06}
  F.~Gross and P.~Agbakpe,
  Phys.\ Rev.\  C {\bf 73}, 015203 (2006)
  [arXiv:nucl-th/0411090].







\bibitem{DeltaDFF2}
  G.~Ramalho, Franz Gross and M.~T.~Pe\~na,
  {\it Electromagnetic form factors of the Delta in a D-wave approach},
  work in preparation



\bibitem{Nozawa90}
  S.~Nozawa and D.~B.~Leinweber,
  Phys.\ Rev.\  D {\bf 42}, 3567 (1990).


\bibitem{Pascalutsa07}
  V.~Pascalutsa, M.~Vanderhaeghen and S.~N.~Yang,
  Phys.\ Rept.\  {\bf 437}, 125 (2007)
  [arXiv:hep-ph/0609004].



\bibitem{Lattice}
  G.~Ramalho and M.~T.~Pe\~na,
  arXiv:0812.0187 [hep-ph], to appear in J.~Phys.~G.




\bibitem{Leinweber01}
  D.~B.~Leinweber, A.~W.~Thomas, K.~Tsushima and S.~V.~Wright,
  Phys.\ Rev.\  D {\bf 64}, 094502 (2001)
  [arXiv:hep-lat/0104013].






\bibitem{Detmold}
  W.~Detmold, D.~B.~Leinweber, W.~Melnitchouk, A.~W.~Thomas and S.~V.~Wright,
  Pramana {\bf 57}, 251 (2001)
  [arXiv:nucl-th/0104043];
  J.~D.~Ashley, D.~B.~Leinweber, A.~W.~Thomas and R.~D.~Young,
  Eur.\ Phys.\ J.\  A {\bf 19}, 9 (2004)
  [arXiv:hep-lat/0308024].


\bibitem{Young}
  R.~D.~Young, D.~B.~Leinweber and A.~W.~Thomas,
  Nucl.\ Phys.\ Proc.\ Suppl.\  {\bf 128}, 227 (2004)
  [arXiv:hep-lat/0311038];
  D.~B.~Leinweber, A.~W.~Thomas, A.~G.~Williams, R.~D.~Young,
J.~M.~Zanotti and J.~B.~Zhang,
  Nucl.\ Phys.\  A {\bf 737}, 177 (2004)
  [arXiv:nucl-th/0308083].





\bibitem{Boinepalli06}
  S.~Boinepalli, D.~B.~Leinweber, A.~G.~Williams,
  J.~M.~Zanotti and J.~B.~Zhang,
  Phys.\ Rev.\  D {\bf 74}, 093005 (2006)
  [arXiv:hep-lat/0604022].





\end{thebibliography}
\end{document}